\documentclass{ansarticle}

\usepackage{anslistings}
\usepackage{amsmath}
\usepackage{amsfonts}
\usepackage{xspace}
\usepackage{a4wide}
\usepackage{comment}

\definecolor{hrefcolor}{rgb}{0.0,0.0,0.8}
\newcommand{\linkcolor}{hrefcolor}
\hypersetup{pdfstartview=FitR}
\hypersetup{pdftitle={The DUNE-Python Module},
            pdfauthor={A. Dedner, M. Nolte},
            breaklinks=true,
            linkbordercolor=\linkcolor,
            urlbordercolor=\linkcolor,
            citebordercolor=\linkcolor,
            runbordercolor=\linkcolor,
            menubordercolor=\linkcolor,
            filebordercolor=\linkcolor,
            runcolor=\linkcolor,
            baseurl={https://gitlab.dune-project.org/staging/dune-python/blob/releases/2.6/},
            extension=
}
\hypersetup{pdfborderstyle={/S/U/W 1}}

\newcommand{\dune}[1][]{\textsc{Dune}\ifx&#1&\else\textsc{-{#1}}\fi\xspace}

\newcommand{\yaspgrid}{\textsc{YaspGrid}\xspace}

\newcommand{\mayavi}{\pyth{Mayavi}\xspace}
\newcommand{\matplotlib}{\pyth{Matplotlib}\xspace}
\newcommand{\quadpy}{\pyth{Quadpy}\xspace}
\newcommand{\numpy}{\pyth{Numpy}\xspace}
\newcommand{\scipy}{\pyth{Scipy}\xspace}
\newcommand{\paraview}{Paraview\xspace}
\newcommand{\vtk}{VTK\xspace}
\newcommand{\cmake}{cmake\xspace}
\newcommand{\pybind}{Pybind11\xspace}
\newcommand{\code}[1]{\lstinline[basicstyle=\small\sffamily]{#1} }
\newcommand{\file}[1]{\code{#1}}

\newcommand{\Atodo}[1]{\todo[inline]{{\bf\color{green}Andreas:} #1}}
\newcommand{\Mtodo}[1]{\todo[inline]{{\bf\color{green}Martin:} #1}}

\newcommand{\pweavecaption}{}
\newcommand{\pweavelabel}{}
\lstnewenvironment{pweavecode}[1][]{\lstset{escapechar=`,moredelim=[il][]{\#tex},style=pythonstyle, title={\codetitlestyle{Python code }},belowcaptionskip=\belowtitleskip, caption=\pweavecaption,label=\pweavelabel}}{}
\lstnewenvironment{pweaveout}[1][]{\lstset{style=gencodestyle, title={\codetitlestyle{Output}}, belowcaptionskip=\belowtitleskip}}{}
\newcommand{\setpweavecaption}[2]{%
  \renewcommand{\pweavecaption}{#1}%
  \renewcommand{\pweavelabel}{#2}%
}

\begin{document}

\title{The \dune[Python] Module}
\author[1]{Andreas Dedner}
\author[2]{Martin Nolte}
\affil[1]{University of Warwick, UK}
\affil[2]{University of Freiburg, Germany}
\runningtitle{The \dune[Python] Module}
\runningauthor{Dedner, Nolte}

\providecommand{\keywords}[1]{\textbf{Keywords: } #1}

\date{July, 2018}

\maketitle


\begin{abstract}
In this paper we present the new \dune[Python] module which provides Python
bindings for the \dune\ core, which is a C++ environment for solving
partial differential equations.
The aim of this new module is to
firstly provide the general infrastructure for exporting realizations of statically polymorphic
interfaces based on just-in-time compilation and secondly to provide bindings for the
central interfaces of the \dune\ core modules. In the first release we
focus on the grid interface.
Our aim is to only introduce a thin layer when
passing objects into Python which can be removed when the object is passed
back into a C++ algorithm. Thus no efficiency is used and little additional code
maintenance cost is incurred. To make the transition for \dune\ users to the
Python environment straightforward the Python classes provide a very
similar interface to their C++ counterparts.
In addition, vectorized versions of many interfaces allow for more efficient
code on the Python side.
The infrastructure for exporting these interfaces, the resulting
for a \dune\ grid are explained in detail in this paper for both experienced
\dune\ users and others interested in a flexible Python environment for
implementing grid based schemes for solving partial differential equations.

\noindent
\end{abstract}

\section{Introduction}

Software for solving complex systems of non linear partial differential
equations (PDEs) have become an important tool for investigating problems in both
academia and industry. PDE problems arise in a wide range of disciplines,
from engineering, physics, biology, medicine, to finance and psychology.
Many different types of schemes are available for solving these problems.
Many of these are grid based discretization methods, e.g., finite
differences, finite volume, and finite element methods. Especially for
problems posed in complex geometries finite volume or finite element
methods are widely used due to their flexibility. For example,
many different types of grids (e.g. partially structured, locally refined) can be combined with
many different finite element spaces, leading to a wide range of
available approaches for solving one and the same problem, each with its
distinct advantage and disadvantage. Especially for users not so familiar
with the mathematical background of these methods, this flexibility can be
difficult to handle. Also the requirements on the underlying data
structures for an efficient realization of these methods make the
implementation of these methods far from easy. This has resulted in
development of a number of open source software packages over the last two
decades, e.g., \cite{alberta:05,dealII81,libMeshPaper,amdis,freefempp,dunepaperII:08}.

The distributed and unified numerics environment (\dune), see \cite{dunepaperI:08},
is based around uniform interfaces for the essential building blocks
of grid based numerical schemes for solving PDEs, e.g., the grid structure,
the finite element spaces, and dense and sparse linear algebra.
Furthermore, it provides a number of realizations of these interfaces.
\dune\ is written in C++, using slim, compile-time static interfaces to
implement numerical algorithms combining both a high level of flexibility
and high efficiency.

Over the last decade, Python has evolved into a frequently used scripting
language in scientific computing. The scripting features of the language
together with tools such as Jupyter notebooks result in a
gentle learning curve for new developers of a package while allowing for
rapid prototyping and easy program control also for experienced users. To
achieve the efficiency required for scientific computing many additional packages are
available which build on libraries written in a programming language such
as Fortran, C, or C++.  Data compatibility is achieved by Pythons buffer protocol (PEP 3118).
Apart from the famous \numpy and SciPy libraries, bindings have been written for
many scientific libraries such as for example PETSc \cite{petsc4py}.

Coupling C++ code with Python is done in many projects and there are
well-established concepts for exporting C++ classes to Python, like
SWIG, Boost-Python, or \pybind (see \cite{pybind11}).
Instead of reinventing the wheel, \dune[Python] builds on \pybind to export
C++ classes and functions to Python.

The first major contribution to solving general systems of PDEs with finite
element methods in Python using a C++ backend is the FEniCS project
\cite{fenics}. A newer development is the Firedrake project \cite{firedrake}. Both
use the domain specific language UFL \cite{ufl} to describe the PDE system
and just-in-time compilation to generate highly optimized code.

A first attempt to provide Python bindings for \dune\ has been undertaken within
the BEM++ project \cite{bempp}.
As Python is an interpreted language, it has been argued that the overhead of
dynamic polymorphism over the static \dune\ interface is negligible.
Therefore, BEM++ only exports a type-erased version of the \dune\ interface
to Python.
While this approach does allow using \dune\ from Python, it only allows
passing the type-erased objects back into other C++ algorithms, replacing
the slim static interface layer by a heavy-weight dynamic one. Also the
realizations of the \dune\ interfaces e.g. the type of the underlying grid structure
is fixed at the configuration stage thus removing part of the flexibility which is a
hallmark of the \dune\ project.

The \dune[Python]\ module described in this paper takes a different approach:
We export the actual \dune\ implementation to Python.
As \dune\ is a C++ template library, the number of classes the user might
possibly instantiate is very large and, in truth, not even finite.
Therefore, we resort to just-in-time (JIT) compilation of any instantiated classes
and algorithms. This project is quite different from the above-mentioned
FEniCS or Firedrake projects in that we aim at providing a
very low level interface with fine granularity for implementing numerical
algorithms directly within Python. We seek to preserve the full flexibility
and modularity of \dune\ so that the numerical schemes can be easily tested
and investigated using the wide range of grid structures available
Finally, by keeping the interface closely aligned with the \dune C++
interface, prototype algorithms written in Python can be easily transferred
to C++, resulting in highly efficient implementations of the algorithms.
Infrastructure for the JIT compilation of functions contained in
single header files makes it straightforward to import the C++ versions of
the algorithms and thus to replace their Python counterparts without
requiring any major code restructuring.
Finally, by exporting the full \dune grid interface this module is well
suited for teaching the basic concepts required for implementing finite
element methods.

Higher level \dune modules like \cite{dune:Fem} can make use
of the concepts presented in this paper for exporting realizations of
static polymorphic C++ interfaces to Python.
In the module \dune[Fempy], which
will be described in a separate paper \cite{dune:Fempy},
we provide higher level functionality for solving PDEs
making use of UFL for the problem formulation. This make solving
these complex problems efficiently straightforward and by combining this with
the interfaces exported in \dune[Python] pre- and post-processing as well
as rapid prototyping of new components can be easily carried out.


This paper is organized as follows.
The mechanisms for generating on-demand Python bindings within the
\dune\ build system are detailed in Section~\ref{sec:PythonBindings}.
We then present a brief overview of Python bindings for the \dune\ core
modules \dune[Common], \dune[Geometry], and \dune[Grid]
in Section~\ref{sec:corebindings}.
Section~\ref{sec:vectorization} then focuses on deviations from the C++
interface to greatly improve performance of the Python code, e.g., through
vectorization.
Finally, we conclude with some numerical experiments in
Section~\ref{sec:examples}.
After some concluding remarks we give some installation instructions in
Appendix~\ref{sec:installation} and a list of interface differences between
Python and C++ in Appendix~\ref{sec:changedinterfaces}.

\section{Python Bindings for Dune Modules}
\label{sec:PythonBindings}

Any \dune environment consists of a set of \dune modules, some of which can
be installed system wide, while others reside in the user's local space.
A \dune module has a name (e.g. \dune[module]) and contains header files
(usually located in a subdirectory \file{dune/module/...}) and possibly source
files compiled into libraries.
Each module can depend on or suggest any number of other \dune modules (as
long as the resulting dependency tree is acyclic).
Usually all \dune modules are then built using a shell script \file{dunecontrol} provided
in the core module \dune[Common] which resolves their interdependencies and builds them
in correct order using \cmake.

\dune contains a number of \emph{core modules} (at the time of writing
these are \dune[Common], \dune[Geometry],
\dune[Grid], \dune[Istl], and \dune[LocalFunctions]).
Their main purpose is to provide interfaces for
the basic building blocks of a numerical scheme, e.g., the required dense
and sparse linear algebra, the underlying grid structure, and for attaching 
the discrete data to the grid. The core modules also contain different
realizations for each of these interfaces; further realizations are available in additional (user) modules.
To obtain a good Python experience, any implementation of an interface, say
the \dune\ grid interface, must export the interface functionality in the
same manner.  This is achieved by defining a set of registration functions
to export this interface to Python and when a specific implementation is
requested, a Python module is build using JIT compilation.

In addition to providing the infrastructure for the JIT compilation
process, the focus of the first release of \dune[Python] is on providing
bindings for \dune[Common], \dune[Geometry],
and \dune[Grid] and these will be described in the following section.
Preliminary bindings are provided for \dune[Istl].

An issue is that additional modules with Python bindings, e.g., a module with an
additional grid implementation like \dune[ALUGrid] or a user
module, are not known during the build process of \dune[Python].
Furthermore, \dune[Python] might be installed into the system so that
it cannot be modified by the user.
Thus \dune[Python] is not suitable for carrying out the JIT compilation
process since it build directory might not be available or writeable
during the running of a Python script and other \dune modules or
third party packages were not available during the configuration phase.
To overcome this problem, a new \dune module (\dune[py]) is generated with a
default location in \file{.cache} within the
user's home directory or within a virtual environment if one is activated.
This module is generated either by the user calling a script in
\file{dune-python/bin} or on the fly when the user imports any part of the
\dune Python package for the first time. More details on both these options
are available on the project's web page or in the \file{README.md} file
available in \dune[Python].

\dune[py] depends on all \dune
modules found in the system or under the location given by the
\file{DUNE_CONTROL_PATH} environment variable. Thus it can depend on both
core and user modules and when it is build the correct search paths are
added. By setting \file{DUNE_CMAKE_FLAGS} non default paths for external
packages (e.g., PETSc or Eigen), compiler flags, etc.\ can be passed to the
build process of \dune[py]. Once this module is set up and configured using
\cmake (using in source build mode)
all files generated on the fly for the JIT process are written to
\file{dune-py/python/dune/generated} which is set up as subpackage of the
\file{dune} namespace package. Each new module that is being generated
is added to the \file{CMakeLists.txt} file and consequently calling \cmake
compiles the module with the include and library paths correctly set to find
all available \dune modules and external packages. With this set up dependency tracking
of source and header files is correctly managed by \cmake.
Modules are only compiled if necessary so that rerunning a Python script
does not incur any compilation cost.

In the following we will discuss how we export realizations of some
interface to Python. We concentrate on statically polymorphic interfaces,
which are sometimes referred to as concepts in C++, since these are
common in many \dune modules.
In general, there is no C++ class defining the interface and, in particular,
there is no abstract base class common to all implementations.
Apart from increasing code maintenance, adding such a base class would greatly
reduce efficiency whenever such a type-erased object is passed into a C++
algorithm.
Instead, we choose to directly export a given realization of a statically
polymorphic interface by defining a common registration function template
filling in the \pybind \cpp{class_} structure.
As an introduction to \pybind is beyond the scope of this paper, we will
assume the reader is familiar with the core concepts of \pybind in the
following; detailed knowledge should not be required though.
In Section~\ref{sec:bindings} we will discuss the general concept of how to provide
bindings for realizations of a static interface located in an example module.
This approach is used for example for the bindings provided in
\dune[Python]. In Section~\ref{sec:registration} the few steps required to export
additional realization located in a further downstream module are detailed,
e.g., an additional grid implementation like \dune[ALUGrid].
Finally in Section~\ref{sec:JIT} we discuss how to compile on demand
single C++ template functions where the template arguments are fixed
according to the type of the parameters passed to the function from within
the Python script.

\subsection{Exporting Statically Polymorphic C++ Class Structure to Python}
\label{sec:bindings}

For the sake of presentation we will consider an example \dune\ module called
\dune[mymodule], which exports the following interface \cpp{Foo}:
\inputcpp{mymodule.hh}{Example interface \cpp{Foo}}{}
In the following we also assume that realizations of this interface
provide a constructor taking a single \cpp{double} parameter. This is not
part of the interface description and we will explain how to export more
general constructors in the following.

To export any implementation of this interface to Python, \dune[MyModule] provides
a file \file{dune/mymodule/py/foo.hh} containing a
function \cpp{registerFoo} with the binding code
\inputcpp{foo.hh}{Registration function}{lstFooExportCode}

Let us assume further that we have a file \file{dune/mymodule/fooimpl.hh} containing
an implementations of this interface, \cpp{FooImplA}:
\inputcpp{fooimpl.hh}{Content of \file{dune/mymodule/fooimpl.hh}}{}

On the Python side we want to provide a function \cpp{fooA} which
constructs an instance of \cpp{FooImplA} with both template and constructor
argument provided dynamically by the user:
\inputpython{__init__.py}{Python binding code}{lstExportingInterfaces}

With this, simply calling \pyth{fooA} will lead to the compilation
of the required python module.
\renewcommand{\pweavecaption}{Requesting the existing interface realization}
\renewcommand{\pweavelabel}{lstCompilingInterfaces}

\begin{pweavecode}
import dune.mymodule as mymodule
fooa = mymodule.fooA(10,dim=2)
print(fooa.foo())
\end{pweavecode}
\begin{pweaveout}
5.0
\end{pweaveout}

The central class used in this example is \pyth{SimpleGenerator}.
Its constructor arguments are the name of the interface to be exported and
the C++ namespace containing the registration function, in this case
\cpp{MyModule::registerFoo}. The \pyth{SimpleGenerator} is a first
implementation of the JIT process and further versions are planned
providing more flexibility.
The \pyth{load} function defined above takes the \pyth{typeName} which is
a string containing the C++ class to be exported. The first argument is a
list of include files to add to the C++ file generated; we will discuss the
other two optional arguments later. This function can be used to export
additional realizations of \cpp{Foo}.

The actual function \pyth{fooA}, when
called by the user, needs to construct the string containing the C++ type
and collect all required include files. To simplify this process we
provide a function \pyth{generateTypeName} described in more
detail in Section~\ref{sec:type registry}
In this case, passing \pyth{a=10} and \pyth{dim=2} returns the values
\pyth{typeName = 'MyModule::FooImplA< 2 >'} and \pyth{includes=[]}.

Next we demonstrate how to export an implementation of \cpp{Foo}
which uses a different constructor and with an additional non interface methods
which we would also like to export:
\inputcpp{foob.hh}{Exporting an extended \cpp{Foo} class}{}
The constructor used previously to export
\cpp{FooImplA} only takes one \cpp{double} so is not suitable
for \cpp{FooImplB} and the \cpp{registerFoo} function will lead
to a compilation error. To avoid this the function can be
overloaded for \cpp{FooImplB}:
\inputcpp{foob.hh}{Specialized registration function for \cpp{FooImplB}}{lstCPPregisterFooB}
An alternative to overloading the registration method would be to hide the
export of the constructor in the original implementation of the
registration using SFINAE, which is beyond the scope of this paper.

Having taken care of the non standard constructor we would also
like to export the additional methods. We could easily do this
by adding the corresponding \pybind method calls to the
specialized registration method. An alternative is to pass an
additional \pyth{Method} instance as argument to the
\pyth{load} function:
\inputpython{foob.py}{Python function to construct the extended class}{lstFooB}
Again constructing an instance of this class is simple
\renewcommand{\pweavecaption}{Instantiating a \cpp{FooBImpl} instance}
\renewcommand{\pweavelabel}{lstFooBInstantiation}

\begin{pweavecode}
from dune.mymodule.foob import fooB
foob = fooB(5,10,dim=2)
print(foob.foo(), foob.aGreaterb())
\end{pweavecode}
\begin{pweaveout}
25.0 False
\end{pweaveout}


\subsection{Type registry}
\label{sec:type registry}

A central concept allowing to export complex interface realizations to
Python is the \emph{type registry}. Each exported class is registered
together with a string containing its full C++ type and a list of the
required includes, i.e., the values of the \pyth{includes} and
\pyth{typeName} arguments passed to the \pyth{load} method.
These values can be retrieved from any instance of the exported Python
class and used as template arguments during the export of another C++ class.
Assume, for example, that a realization of our \cpp{Foo} interface is based
on some other realization of \cpp{Foo}:
\inputcpp{fooc.hh}{Exporting a meta implementation}{}
We would now like to construct \cpp{FooC<FooA>} given only an instance of
\pyth{FooA} by passing this instance as parameter to a \pyth{fooC} method,
i.e., by calling \pyth{fooc = fooC(fooa)}:
\inputpython{fooc.py}{Python function to construct the meta implementation}{lstFooC}
Here, the type registry allows us to extract the required class name and
include files from the instance \pyth{fooa}, i.e.,
\pyth{generateTypeName} returns
\pyth{'MyModule::FooImplC< MyModule::FooImplA< 2 > >'} and
\pyth{['dune/mymodule/fooimpl.hh', 'dune/mymodule/py/foo.hh']}.
The code generated when calling \pyth{fooc = fooC(fooa)} is
\begin{c++}{Code generated to export the implementation of \cpp{FooImplC< FooImplA >}}{lst:exportCPP}
#include <config.h>
#define USING_DUNE_PYTHON 1

#include <dune/mymodule/fooimpl.hh>
#include <dune/mymodule/py/foo.hh>
#include <dune/mymodule/fooc.hh>
#include <dune/mymodule/py/fooc.hh>

#include <dune/python/common/typeregistry.hh>
#include <dune/python/pybind11/pybind11.h>
#include <dune/python/pybind11/stl.h>

typedef MyModule::FooImplC< MyModule::FooImplA< 2 > > DuneType;

PYBIND11_MODULE( MyModule_Foo_c348917f6caddd66354a05b688c1a4c3, module )
{
  using pybind11::operator""_a;
  auto cls = Dune::Python::insertClass< DuneType >( module, "Foo",
    Dune::Python::GenerateTypeName(
      "MyModule::FooImplC< MyModule::FooImplA< 2 > >"),
    Dune::Python::IncludeFiles{
      "dune/mymodule/fooimpl.hh","dune/mymodule/py/foo.hh",
      "dune/mymodule/fooc.hh","dune/mymodule/py/fooc.hh"}).first;
  MyModule::registerFoo( module, cls );
}
\end{c++}
The C++ type of the interface realization is provided as \cpp{DuneType}.
Note the \cpp{USING_DUNE_PYTHON} preprocessor variable, which can be
used enable code specific to the python bindings in the included header files.
Thus, header files can be easily used both in a C++ context and in a Python context.
The correct entry in the type registry is guaranteed by using the function
\cpp{insertClass}, which takes the \pybind module, the Python name for the
class, the type name, and the include's file names.
The return is a pair containing the \pybind \cpp{class_} instance and a \cpp{bool}
indicating whether the class was newly inserted into in the type registry.
More details on the type registry can be found in the code documentation.

\subsection{Providing Bindings for Additional Modules}
\label{sec:registration}

Now, let us assume that a further realization \cpp{FooImplE}
of our \cpp{FooInterface} was provided by another \dune module
\dune[somebodyelsesmodule] (which we will abbreviate
with \emph{sem} in the following). So the class \cpp{FooImplE}
is now in the namespace \cpp{Sem} and defined in
the header file \file{dune/sem/fooe.hh}. To add the Python bindings the
authors of \dune[sem] have to carry out the following steps:
\begin{enumerate}
  \item
    First suggest \file{dune-python} in the \file{dune.module} file and add the
    following to the main \file{CMakeLists.txt}:
    \begin{anycode}{}{}
if(dune-python_FOUND)
  add_subdirectory(python)
  dune_python_install_package(PATH "python")
endif()
    \end{anycode}

  \item
    Add a new subdirectory \file{python} containing a \file{setup.py.in}
    \inputpython{setup.py.in}{}{}
    and a \file{CMakeLists.txt} containing the following two lines:
    \inputanycode{CMakeListsp.txt}{}{}

  \item
    Add a new subdirectory \file{python/dune} containing a
    \file{__init__.py} declaring the \pyth{dune} namespace
    \inputpython{__init__.py}{}{}
    together with a \file{CMakeLists.txt} containing the lines
    \inputanycode{CMakeListspd.txt}{}{}

  \item
    Finally add a new subdirectory \file{python/dune/sem} containing the
    actual registration function \pyth{fooE} in \file{__init__.py}:
    \inputpython{__init__.py}{}{}
    Again, add a \file{CMakeLists.txt} containing
    \inputanycode{CMakeListspds.txt}{}{}
\end{enumerate}
Note that we import the \pyth{load} function defined in the \pyth{dune.mymodule}
subpackage (see Listing~\ref{lstExportingInterfaces}).

Now importing \pyth{dune.sem} and executing \pyth{foo = dune.sem.fooE(10)}
will, if required, build the Python module and construct an instance of \cpp{FooE}.

\subsection{Just-in-Time Compilation of Single Functions}
\label{sec:JIT}

In addition to the infrastructure for exporting realizations of static
interfaces to Python, the \pyth{dune.generator} module also provides
simples function to compile and invoke C++ functions.
Compilation is, again, carried out within the \dune[py] module, so all
\dune modules and their configuration are available.
Types of parameters are automatically deduced at runtime allowing, again,
for a very generic style of programming. In fact this approach makes it
easy to use functions both within the \dune[Python] framework and within
a purely C++ based \dune application.

Let's assume we have implemented a
function template \cpp{bar} in a header file \file{bar.hh}:
\inputcpp{bar.hh}{C++ header with \cpp{bar} function}{lstbarCPPHeader}
Now, we want to call this function from our Python script passing instances of
realizations of \cpp{FooInterface} as arguments.
\renewcommand{\pweavecaption}{Calling \cpp{bar} using JIT compilation}
\renewcommand{\pweavelabel}{lstPyBar1}

\begin{pweavecode}
from dune.generator import algorithm
from dune.mymodule import fooA
from dune.mymodule.fooc import fooC
foo_a = fooA(10, dim=2)
foo_c = fooC(foo_a)
ret = 0
for p in range(10):
  ret += algorithm.run('bar', 'bar.hh', foo_c, foo_a, p+1)
print(ret)
\end{pweavecode}
\begin{pweaveout}
80.61733715087786
\end{pweaveout}

This needs to generate bindings for the function \cpp{bar} using
\cpp{FooImplC< FooImplA >} and \cpp{FooImplA< 2 >} as template
arguments.
During the actual JIT compilation, very little binding code needs to be
generated as shown below:
\begin{c++}{Generated code to produce the Python binding}{lstBindingCPP}
#include <config.h>
#define USING_DUNE_PYTHON 1

#include <cmath>
template <class FooImpl1, class FooImpl2>
double bar(FooImpl1 &foo1, FooImpl2 &foo2, int p)
{
  return std::pow(foo1.foo(),foo2.foo()/double(p));
}

#include <dune/mymodule/fooimpl.hh>
#include <dune/mymodule/py/foo.hh>
#include <dune/mymodule/fooc.hh>

#include <dune/python/common/typeregistry.hh>
#include <dune/python/pybind11/pybind11.h>

PYBIND11_MODULE(
  algorithm_8c792602628758c06c506b5c15532df3_cbab2c6e7eb87a8684354bcab72ed606,
  module )
{
  module.def( "run", [] ( MyModule::FooImplC< MyModule::FooImplA< 2 > > & arg0, MyModule::FooImplA< 2 > & arg1, int arg2 ) {
      return bar( arg0, arg1, arg2 );
    } );
}
\end{c++}
Note how the correct include files and types have been extracted from the
arguments passed to \pyth{algorithm.run}, again using the type registry
described in Section~\ref{sec:type registry}.
Although the module will only be loaded once, repeated calls to
\pyth{algorithm.run}, e.g., in the loop shown above, will incur some
unnecessary costs.
Therefore, the actual generation and loading of the module can be separated
from the execution of the function:
\renewcommand{\pweavecaption}%
{Calling \cpp{bar} using JIT compilation}
\renewcommand{\pweavelabel}{lstPyBar2}

\begin{pweavecode}
ret = 0
bar = algorithm.load('bar', 'bar.hh', foo_c, foo_a, 0)
for p in range(10):
  ret += bar(foo_c, foo_a, p+1)
print(ret)
\end{pweavecode}
\begin{pweaveout}
80.61733715087786
\end{pweaveout}

Note that we still pass the arguments to the \pyth{load} function to deduce
C++ argument types.
However, the actual instances upon calling the function \pyth{bar} need not
coincide with the instances passed to the \pyth{load} method.

\section{Python Bindings for the Core \dune Modules}
\label{sec:corebindings}

In this section we show how the approach for exporting statically polymorphic
interfaces presented in Section~\ref{sec:PythonBindings} is used to provide bindings for the
core \dune interfaces. To this end we discuss the central components for
implementing grid based numerical scheme in Python based on \dune[Python].
In Section \ref{sec:gridconstruction} we discuss grid construction,
followed by a discussion of the central parts of the grid interface in
Section~\ref{sec:basicgridusage}.
The central concept of defining functions over a given grid
is discussed in Section \ref{sec:gridfunctions}, which are functions that
can be evaluated on a given element of the grid. We present both the
\emph{bind/unbind} approach used on the C++ side, as well as
a more direct evaluation method. Grid functions can be easily constructed
using decorators. A central part of any numerical scheme
is the construction of discrete grid function requiring to attach degrees
of freedom to different parts of the grid; this is discussed in Section
\ref{sec:attachingdata}.
While visualization of the grid and of grid functions  using
\matplotlib is used throughout this section, more details on output,
e.g., to \vtk files is discussed in Section~\ref{sec:output}.
We conclude this section by discussing parallelization support.

A full Python script containing all the example code shown in the following
can be found in the \dune[Python] module in the folder \file{doc/paper}.

The central classes we want to describe are contained in the \pyth{dune.grid} module.
The \dune[Grid] module depends on \dune[Common] and \dune[Geometry]. The
first contains some dense matrix-vector routines and some other helper
classes, e.g., for parallelization support. \dune[Geometry] contains the reference
elements and quadrature rules required, for example, to implement in finite
element methods.

\subsection{A quick survey of \dune[Common] and \dune[Geometry]}
\label{sec:CommonAndGeometry}


In the following we will only describe the
corresponding Python modules \pyth{dune.common} and \pyth{dune.geometry} without
going into detail.

The python module \pyth{dune.common} provides access to the
\pyth{FieldVector} and \pyth{FieldMatrix} classes:
\renewcommand{\pweavecaption}%
{Instantiating a \dune \cpp{FieldVector} defined in \dune[Common]}
\renewcommand{\pweavelabel}{lstDuneCommon}

\begin{pweavecode}
from dune.common import FieldVector, FieldMatrix
x = FieldVector([0.25,0.25,0.25])
\end{pweavecode}

These dense vectors and matrices are used in many places within the dune
interface, especially for geometrical representations of grid elements.
In general, a list or tuple of correct length can be passed to any function or
method expecting a \cpp{FieldVector} in the \dune C++ interface.

The \pyth{dune.geometry} module provides a class representing a \dune reference
element. It is constructed from a \emph{geometry type}, which encodes the
dimension and a basic shape, e.g., simplex or (hyper)cube.
The following code snippet shows how to obtain the reference element for a
two-dimensional simplex and print all its corners.
\renewcommand{\pweavecaption}%
{Reference elements for a 2d triangle}
\renewcommand{\pweavelabel}{lstDuneGeometry}

\begin{pweavecode}
import dune.geometry
geometryType = dune.geometry.simplex(2)
referenceElement = dune.geometry.referenceElement(geometryType)
print("\t".join(str(c) for c in referenceElement.corners))
\end{pweavecode}
\begin{pweaveout}
(0.000000, 0.000000)    (1.000000, 0.000000)    (0.000000, 1.000000)
\end{pweaveout}

Note that, for convenience, we could have used \pyth{triangle} instead of
\pyth{simplex(2)}.
Experienced \dune users will also note a slight deviation for the C++ interface
in the above code snippet.
The Python property \pyth{corners} returns a tuple of corners while the C++ method
with the same name returns the number of corners.
We will go into
more detail in the next section (or see Appendix~\ref{sec:changedinterfaces}).

Similarly, it is straightforward to write a quadrature loop for the reference
element of a given geometry type:
\renewcommand{\pweavecaption}%
{Print quadrature rule for a triangle}
\renewcommand{\pweavelabel}{lstDuneQuadRule}

\begin{pweavecode}
for p in dune.geometry.quadratureRule(geometryType, 3):
    print(p.position, p.weight)
\end{pweavecode}
\begin{pweaveout}
(0.333333, 0.333333) -0.28125
(0.600000, 0.200000) 0.2604166666666667
(0.200000, 0.600000) 0.2604166666666667
(0.200000, 0.200000) 0.2604166666666667
\end{pweaveout}

\subsection{Grid Construction}
\label{sec:gridconstruction}

The \dune[Grid] interface can be implemented in various ways and there is hardly
a common denominator to the data required to construct a grid.
Therefore, \dune traditionally does not require any specific constructor for a
grid implementation.
However, basic data formats are suitable to construct various grid
implementations.
For example, most grid implementations are able to represent a Cartesian grid
and most unstructured grid implementations can be constructed from a Gmsh file,
given support for the used geometry types.
On the C++ side, \dune provides rather complicated factory concepts to support
different construction mechanisms in addition to custom grid constructors.

To unify the grid constructors, \dune[Python] uses Python's dynamic type system
and requires constructs a grid from an abstract \emph{domain} description.
Conceptually, this can be anything from a file, e.g., in the \dune grid
format (DGF) or the Gmsh format, to a highly specialized data structure
supported by exactly one grid implementation.

The simplest such domain is the \pyth{cartesianDomain}, which can be used to
construct a 2d grid as follows:
\renewcommand{\pweavecaption}{Grid construction of a Cartian grid using the \yaspgrid implementation}
\renewcommand{\pweavelabel}{lstYaspGridConstruction}

\begin{pweavecode}
from dune.grid import cartesianDomain, yaspGrid
domain = cartesianDomain([0, 0], [1, 0.25], [15, 4])
yaspView = yaspGrid(domain)
\end{pweavecode}

This constructs a new \yaspgrid, which is an efficient Cartesian grid implementation
provided by \dune[Grid], on the domain $[0,1]\times[0,0.25]$,
subdivided into $15 \times 4$ cells.
For convenience, the same result can be achieved by
\pyth{dune.grid.structuredGrid([0,0],[1,0.25],[15,4])}, if any structured grid
implementation will do.


In the C++ interface, a \dune grid is hierarchical.
Refining a grid globally or locally does not result in a new grid but adds
child elements to the grid hierarchy.
Frequently, only the leaf level of the hierarchy is used in numerical
computations, e.g., using a finite element method.
For convenience, the \dune[Python] grid construction functions always return
this leaf view of the underlying hierarchical grid, i.e., iterating over the
elements of a refined grid will return the finest elements in the grid.
As a view, however, this object cannot be modified directly and the hierarchical
grid has to be modified instead..
For example, globally refining the grid is done as follows:
\renewcommand{\pweavecaption}{Globally refining a hierarchical grid given its leaf view \pyth{grid}}
\renewcommand{\pweavelabel}{lstGlobalRefine}

\begin{pweavecode}
yaspView.plot()
yaspView.hierarchicalGrid.globalRefine()
yaspView.plot()
\end{pweavecode}
\begin{figure}[htpb]
\center
\includegraphics[width= 0.475\textwidth]{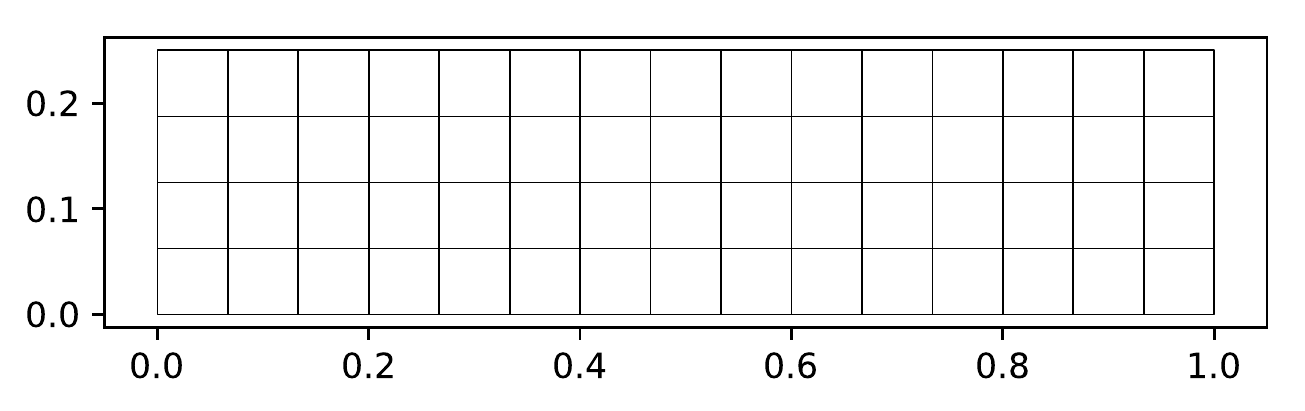}
\includegraphics[width= 0.475\textwidth]{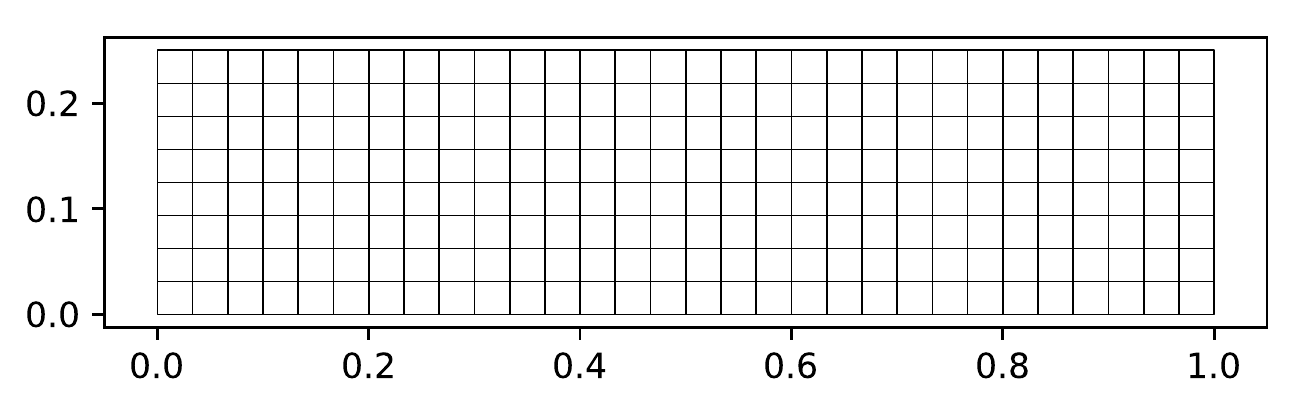}
\caption{refinement of a cartesian grid}
\label{fig:AccessingHGrid}
\end{figure}

To visualize the leaf view, we have used a utility method \pyth{plot}.
This is a convenience extension to the \dune \cpp{GridView} interface,
plotting the grid using \matplotlib.
Note how modifying the hierarchical grid affected its leaf view.

The way \pyth{globalRefine} modifies the grid depends on the grid
implementation.
For example, the following snippet instantiates a triangular grid provided by
the \dune[ALUGrid] module \cite{dune:alugrid} by subdividing each square into
two triangles.
Upon global refinement, this grid implementation uses bisection to conformingly
split each triangle in two.
\renewcommand{\pweavecaption}{Constructing a triangular grid with bisection refinement on $[0,0.25]\times[0,1]$}
\renewcommand{\pweavelabel}{lstALUConformGridConstruction}

\begin{pweavecode}
from dune.alugrid import aluConformGrid
aluView = aluConformGrid(domain)
aluView.plot()
aluView.hierarchicalGrid.globalRefine()
aluView.plot()
\end{pweavecode}
\begin{figure}[htpb]
\center
\includegraphics[width= 0.475\textwidth]{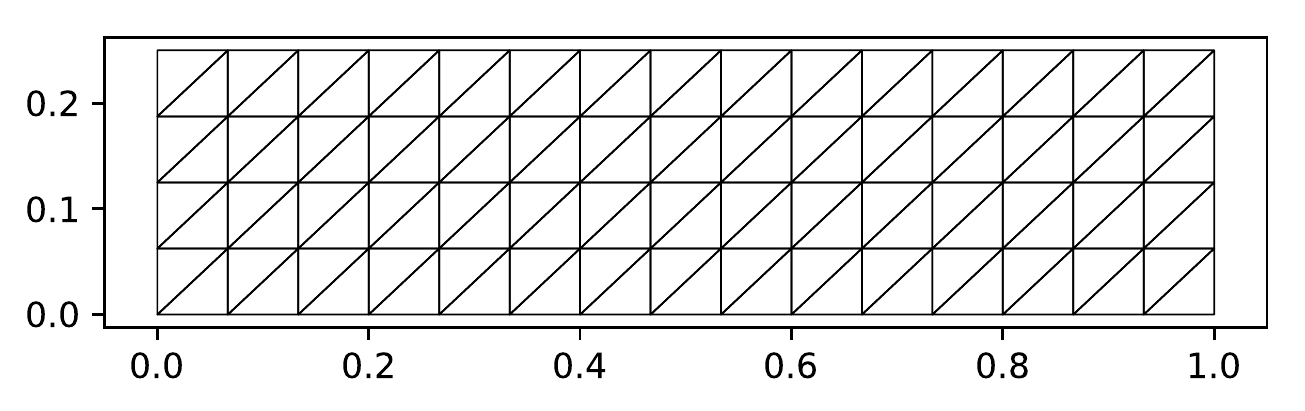}
\includegraphics[width= 0.475\textwidth]{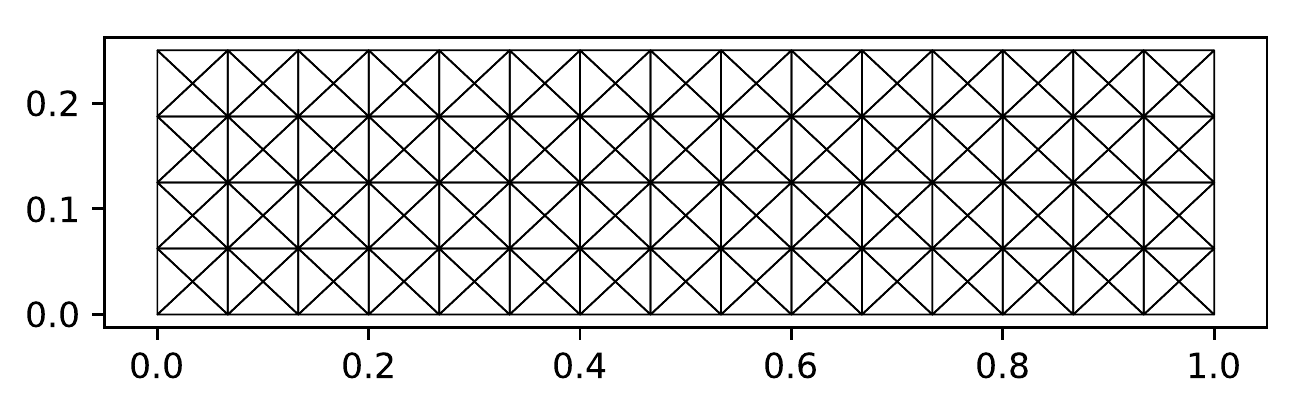}
\caption{Refinement of a simplex grid using bisection}
\label{fig:SimplexGridConstruction}
\end{figure}

Notice that, although \dune[ALUGrid] is not part of the \dune core but exports
Python bindings on its own, this only differes from the construction of a
\cpp{YaspGrid} in the name of the construction function.

The simplest way to set up an unstructured grid is by passing in the vertex
coordinates and the vertex numbers for each element in the grid:
\renewcommand{\pweavecaption}%
{Setting up an unstructed triangular grid}
\renewcommand{\pweavelabel}{lstConstructUnstructuredGrid}

\begin{pweavecode}
vertices = [(0,0), (1,0), (1,0.6), (0,0.6), (-1,0.6), (-1,0), (-1,-0.6), (0,-0.6)]
triangles = [(2,0,1), (0,2,3), (4,0,3), (0,4,5), (6,0,5), (0,6,7)]
aluView = aluConformGrid({"vertices": vertices, "simplices": triangles})
aluView.plot(figsize=(5,5))
aluView.hierarchicalGrid.globalRefine(2)
aluView.plot(figsize=(5,5))
\end{pweavecode}
\begin{figure}[htpb]
\center
\includegraphics[width= 0.4\textwidth]{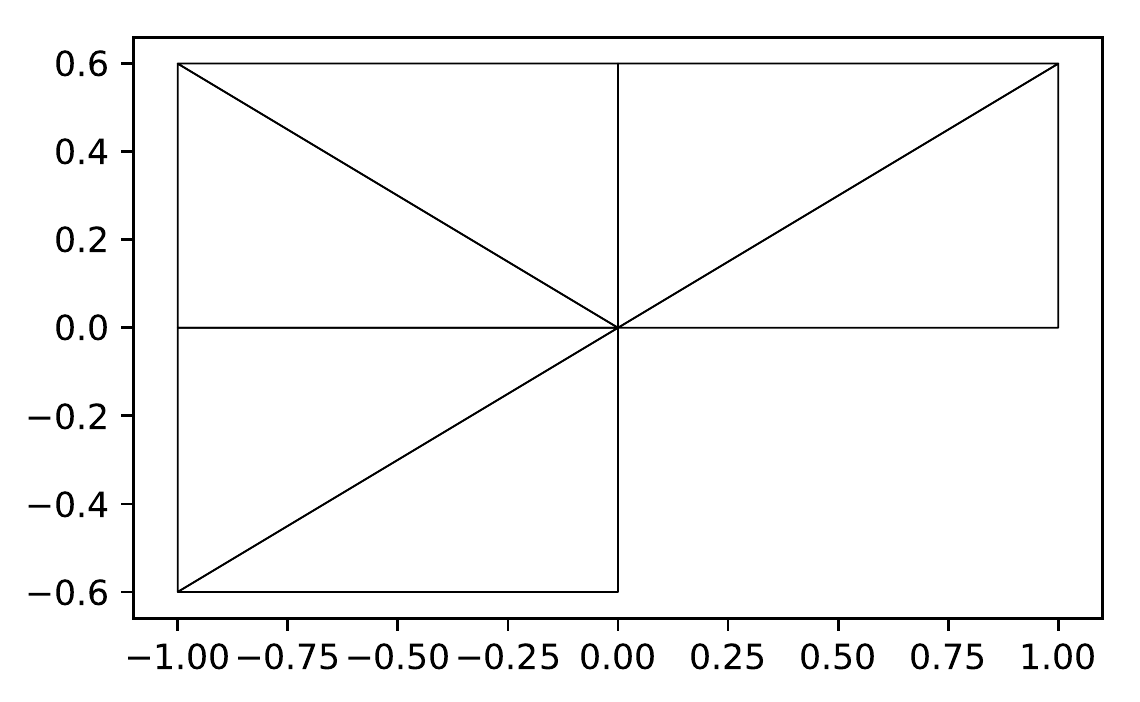}
\includegraphics[width= 0.4\textwidth]{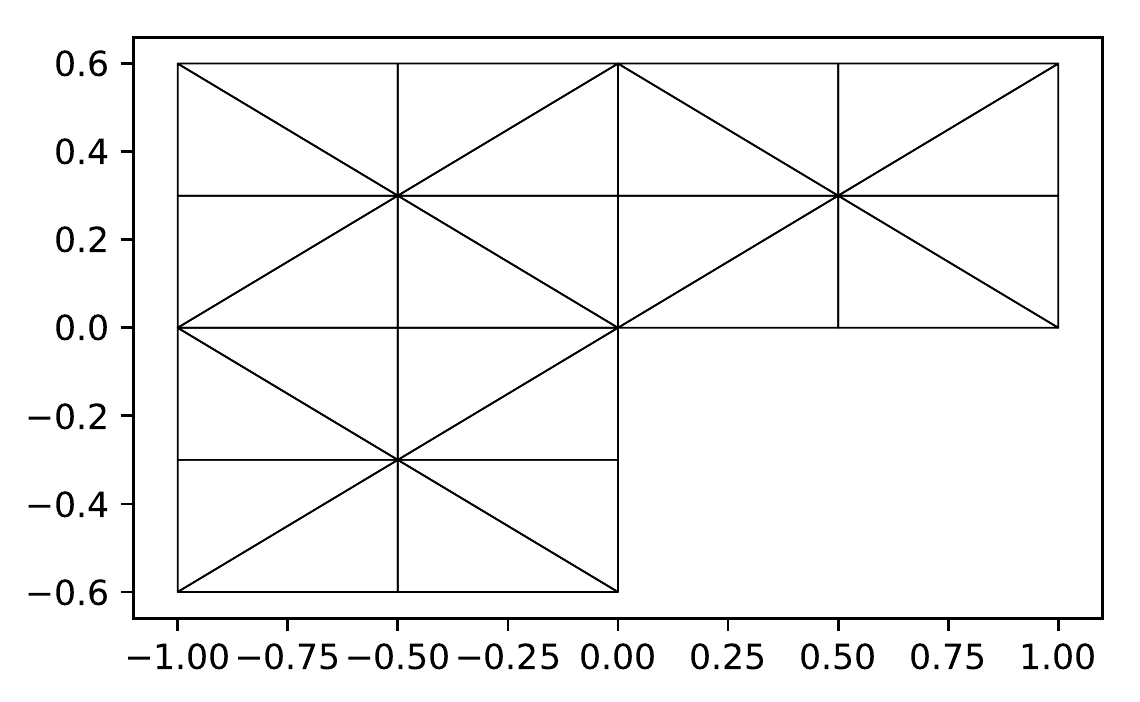}
\caption{An unstructed grid before (left) and after (right) refinement}
\label{fig:ConstructUnstructuredGrid}
\end{figure}

While we use lists here, any iterable structure is possible.
However, for performance reasons we recommend the use of \numpy arrays
for larger grids.

So far, we have been refining every element in our constructed grids.
When preparing a grid, we might also want to refine it locally.
Let us, for example, consider the above grid and focus refinement around the
corner at the origin:

\renewcommand{\pweavecaption}{Local refining around the origin}
\renewcommand{\pweavelabel}{lstLocalRefinement}

\begin{pweavecode}
from dune.grid import Marker
aluView.plot(figsize=(5,5))
for i in range(1,5):
    def mark(e):
        x = e.geometry.center
        return Marker.refine if x.two_norm < 0.64**i else Marker.keep
    aluView.hierarchicalGrid.adapt(mark)
    aluView.plot(figsize=(5,5))
\end{pweavecode}
\begin{figure}[htpb]
\center
\includegraphics[width= 0.195\linewidth]{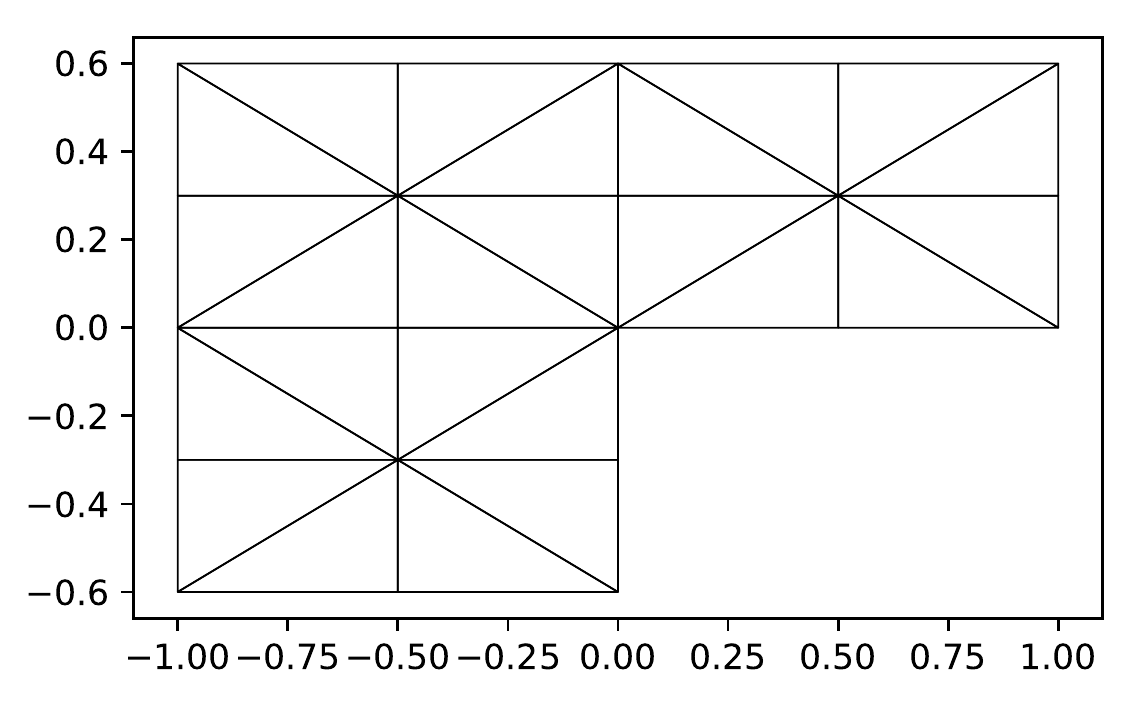}
\includegraphics[width= 0.195\linewidth]{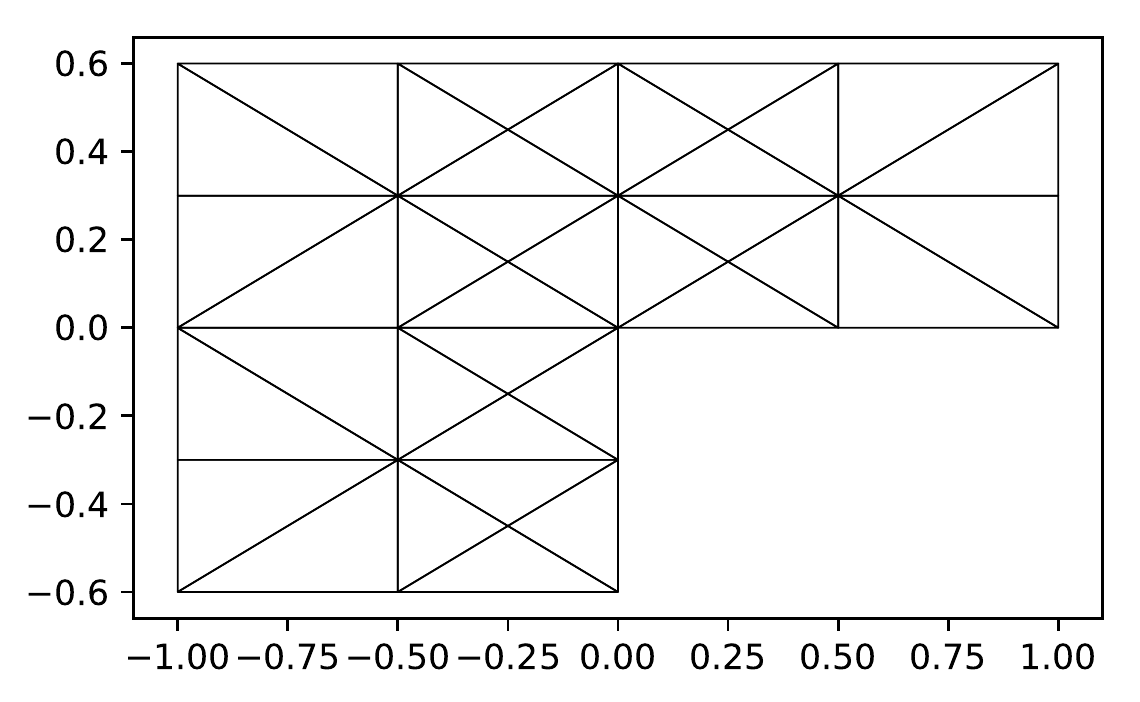}
\includegraphics[width= 0.195\linewidth]{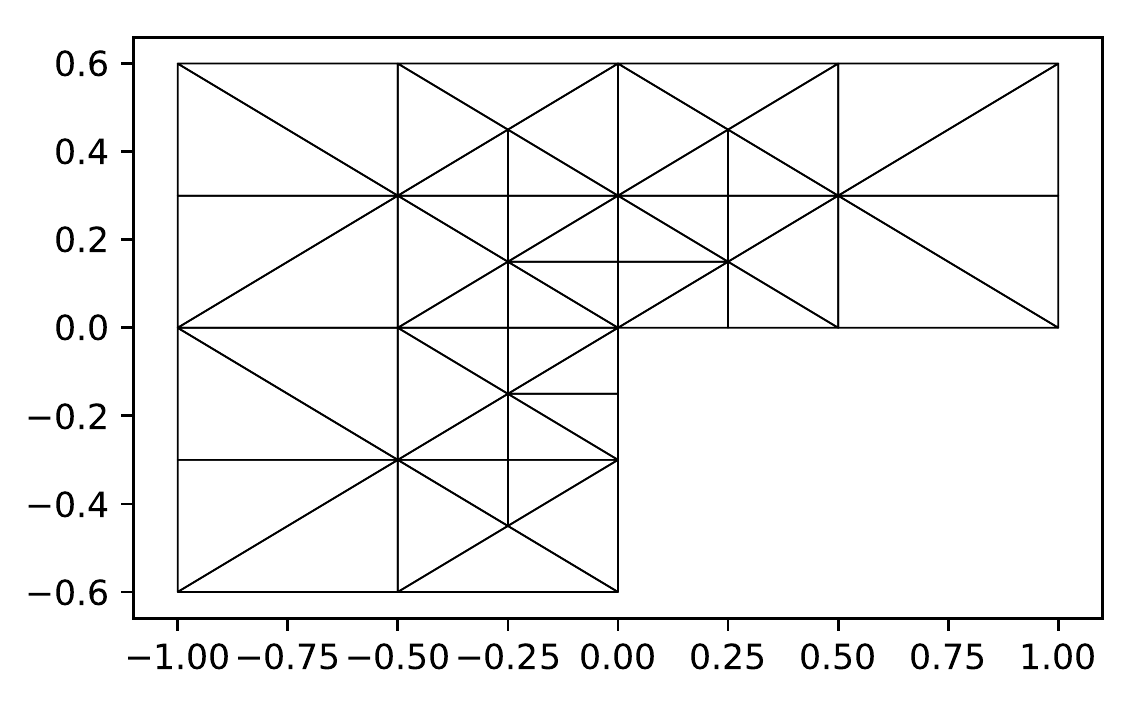}
\includegraphics[width= 0.195\linewidth]{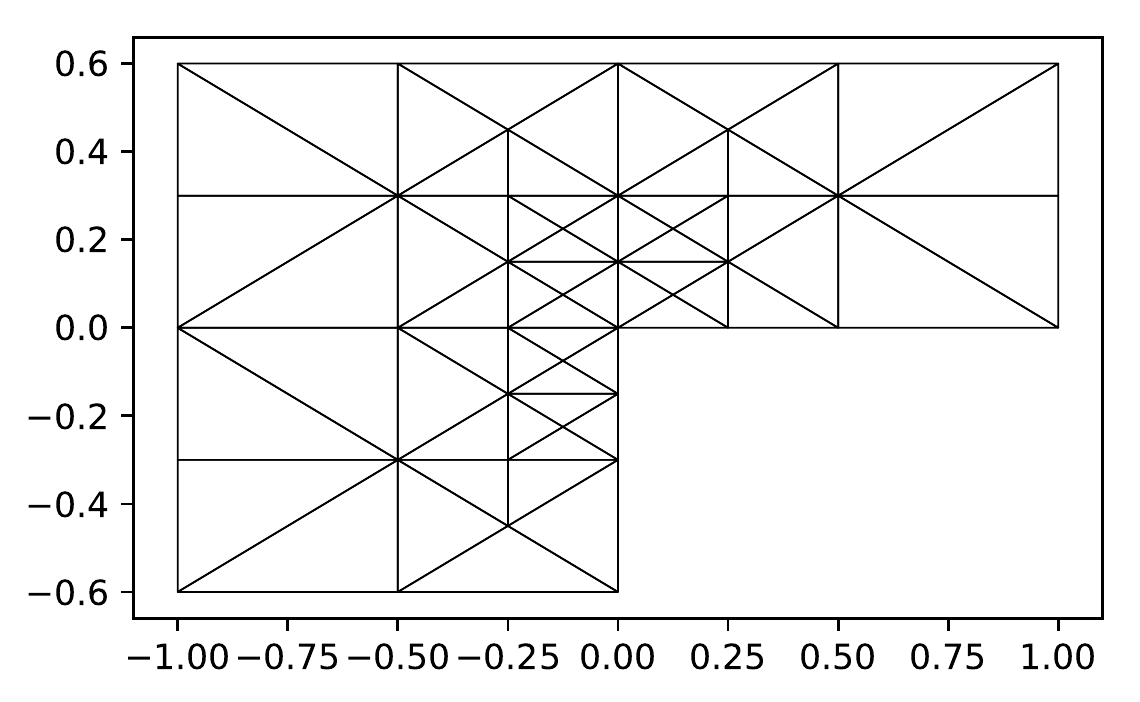}
\includegraphics[width= 0.195\linewidth]{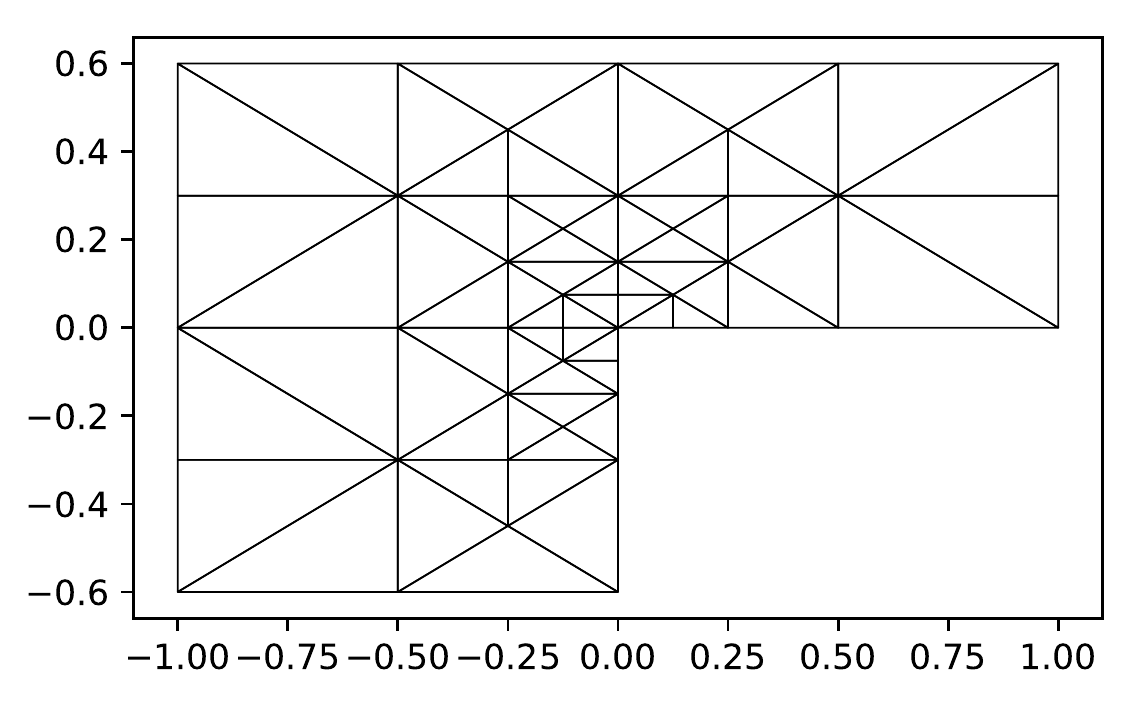}
\caption{Local refined grid}
\label{fig:LocalRefinement}
\end{figure}

Here, we used the \pyth{geometry} property of the element to determine
the distance of its barycenter to the origin.
Obtaining geometrical and topological information from the grid is described
in detail in Section~\ref{sec:basicgridusage}.

Let us assume we want to use this conforming grid as a macro grid to a triangular
grid quartering each element on refinement, e.g., \pyth{aluSimplexGrid}.
To do so, we simply obtain the \pyth{vertices} and \pyth{triangles} arrays
from our grid implementation, e.g.,
\renewcommand{\pweavecaption}{Creating a grid from a leaf view}
\renewcommand{\pweavelabel}{lstCreateFromGrid}

\begin{pweavecode}
from dune.alugrid import aluSimplexGrid
vertices = aluView.coordinates()
triangles = [aluView.indexSet.subIndices(e, 2) for e in aluView.elements]
aluView = aluSimplexGrid({"vertices": vertices, "simplices": triangles})
\end{pweavecode}

Here, \pyth{coordinates} is another convenience function returning the coordinates
of all vertices as a two-dimensional \numpy array.
Of course, we could also have manipulated the coordinates before passing
them on to \pyth{aluSimplexGrid}.
The \pyth{indexSet} property will be described in more detail in Section
\ref{sec:attachingdata}.

All steps taken to construct our triangular grid could also have been done in
C++.
However, the \dune[Python] convenience methods and direct plotting through
\matplotlib make the manual construction of desired macro grids much more
efficient.
After all, dumping the grid into, e.g., a DGF file, we could still load it from
C++.

\subsection{Basic Grid Usage}
\label{sec:basicgridusage}

Now that we know how to construct a grid, let's see what we can do with it.
For the sake of presentation, let us consider a tessellation of the unit square
into two triangles:
\renewcommand{\pweavecaption}{Triangulation of the unit square}
\renewcommand{\pweavelabel}{lstUnitSquare}

\begin{pweavecode}
vertices = [(0,0), (1,0), (1,1), (0,1)]
triangles = [(2,0,1), (0,2,3)]
unitSquare = aluSimplexGrid({"vertices": vertices, "simplices": triangles})
print(unitSquare.size(0),"elements and",unitSquare.size(2),"vertices")
\end{pweavecode}
\begin{pweaveout}
2 elements and 4 vertices
\end{pweaveout}

A \dune grid can be considered as a container of \emph{entities} together with
their relation to each other.
Here, the term entity collectively refers to any topological object in the grid,
e.g., a vertex, an edge, facet, or an element.
Entities are differentiated by their codimension w.r.t.\ the grid dimension, i.e.,
the dimension of its elements.
For example, in a tetrahedral grid, the codimension of a facet is 1 and the
codimension of a vertex is 3. So in the above code
\pyth{unitSquare.size(0)} returns the number of entities of codimension
zero, i.e., the number of triangles.
We can iterate over all entities in the (leaf) grid as follows:
\renewcommand{\pweavecaption}{Iterating over all entities in a grid}
\renewcommand{\pweavelabel}{lstGridIterateAllEntities}

\begin{pweavecode}
for codim in range(0, unitSquare.dimension+1):
    for entity in unitSquare.entities(codim):
        print(", ".join(str(c) for c in entity.geometry.corners))
\end{pweavecode}
\begin{pweaveout}
(1.000000, 1.000000), (1.000000, 0.000000), (0.000000, 0.000000)
(0.000000, 0.000000), (0.000000, 1.000000), (1.000000, 1.000000)
(0.000000, 0.000000), (1.000000, 0.000000)
(0.000000, 0.000000), (1.000000, 1.000000)
(0.000000, 0.000000), (0.000000, 1.000000)
(1.000000, 0.000000), (1.000000, 1.000000)
(1.000000, 1.000000), (0.000000, 1.000000)
(0.000000, 0.000000)
(1.000000, 0.000000)
(1.000000, 1.000000)
(0.000000, 1.000000)
\end{pweaveout}

As an example, we have printed the position of the corner for each entity.
Geometrical information on entities will be discussed later in this section.

In general, we define \emph{elements} to be entities of codimension 0,
\emph{facets} to be entities of codimension 1, \emph{edges} to be
entities of dimension 1, and \emph{vertices} to be entities of dimension 0.
Using this notation, we can also perform the iteration over all edges as follows:
\renewcommand{\pweavecaption}{Iterating over all edges in a grid}
\renewcommand{\pweavelabel}{lstGridIterateAllEdges}

\begin{pweavecode}
for edge in unitSquare.edges:
    print(", ".join(str(c) for c in edge.geometry.corners))
\end{pweavecode}

Just for comparison the following listing shows the same loop based on the C++
interface:
\begin{c++}{Iterating over all edges in a grid using the \dune C++ interface}{lstGridIterateEdgesCPP}
for( edge : edges( unitSquare ) )
{
  const auto geo = edge.geometry();
  std::cout << geo.corner( 0 ) << ", " << geo.corner( 1 ) << std::endl;
}
\end{c++}

In \dune, entities are view-only objects providing the properties \pyth{geometry},
\pyth{level}, \pyth{type}, and \pyth{partitionType}.
For the Python interface, we added the property \pyth{referenceElement}, returning the
reference element of the entity. The same method is also available on the
\pyth{Geometry} classes.

The \pyth{geometry} property models the geometric realization of the entity, i.e.,
a mapping from the local coordinates within the reference element to the global
coordinates in Euclidean space.
The global coordinates of a local point are returned by the method
\pyth{toGlobal}; the local coordinates of a global point can be obtained
via \pyth{toLocal}.
To transform tangential vectors, the Jacobian of the reference mapping and its
inverse are provided by the methods \pyth{jacobianTransposed} and
\pyth{jacobianInverseTransposed}, which are constant if the property \pyth{affine}
is \pyth{True}.
In addition, some properties of the mapping's image are exported: \pyth{center},
\pyth{corners}, and \pyth{volume}.



\subsection{Grid Functions}
\label{sec:gridfunctions}

Having a computational grid, we can define functions on it, e.g., a piecewise
constant function.
To evaluate a piecewise constant function in an arbitrary point $x$, we
would first need to find the element $T$ containing $x$.
However, in practice we usually derive $x$ from a given $T$ and a local position
$\hat x$ in the reference element of $T$, i.e., $T$ is in most cases already available.

Many \dune modules therefore use the concept of \emph{localizable} or
\emph{bindable} grid functions.
These functions are associated with a grid and can be localized to an element
in some manner; this localization is usually referred to as a
\emph{local function}. In the following we will use the term
\emph{grid function} to refer to this specific class of functions.

On the C++ side, \dune[Python] adopts the following interface: For any grid
function, there must be a free-standing function \cpp{localFunction}, which
must be accessible by argument-dependent lookup, to construct a localizable
view of the function.
This view can then be bound to an element and evaluated in a local position
$\hat x$.
An example use could look as follows:
\begin{c++}{}{}
auto lf = localFunction( gridFunction );
lf.bind( element );
auto y = lf( hatx );
lf.unbind();
\end{c++}
The same concept is also accepted by the \cpp{VTKWriter} class available in
\dune[Grid].

We replicate this concept in Python as closely as possible.
As Python does not have the concept of an argument-dependent lookup, we expect
the grid function to provide a method \pyth{localFunction} instead:
\begin{python}
lf = gridFunction.localFunction()
lf.bind(element)
y = lf(hatx)
lf.unbind()
\end{python}

As mentioned above this approach follows the C++ interface as closely as
possible. It is especially useful in the case that multiple evaluations on
one element are required and the \pyth{bind} method
is expensive, e.g., because local degree of freedoms have to be retrieved
from a global vector for a finite element function. For single evaluations
or non time critical code the approach given above is cumbersome and leads
to a high number of inefficient calls between Python and C++. Thus we
provide a simplified interface to evaluate grid functions:
\begin{python}
gridFunction(element, hatx)
\end{python}

Any function on the grid's domain can be easily turned into a grid
function by using the geometric mapping from the reference element.
\dune[Python] provide a \pyth{gridFunction} decorator to add the
\pyth{localFunction} method to a given function:
\renewcommand{\pweavecaption}{Decorating a globally defined function as a grid function}
\renewcommand{\pweavelabel}{lstGF1}

\begin{pweavecode}
@dune.grid.gridFunction(aluView)
def f(x):
    return math.cos(2.*math.pi/(0.3+abs(x[0]*x[1])))
\end{pweavecode}

On the other hand, we might want to implement a piecewise function, i.e., a
function $f(T, \hat x)$ and use it as a grid function.
This works in the same manner:
\renewcommand{\pweavecaption}{Decorating a locally defined function as a grid function}
\renewcommand{\pweavelabel}{lstGF2}

\begin{pweavecode}
@dune.grid.gridFunction(aluView)
def hat0(element,hatx):
    return 1-hatx[0]-hatx[1]
\end{pweavecode}

Now we can, for example, compute the maximum value of \pyth{f} at
the barycenter of all elements using the extended grid function interface
for direct evaluation provided by the decorator:
\renewcommand{\pweavecaption}%
{Using a direct evaluate method on grid function objects}
\renewcommand{\pweavelabel}{lstGF4}

\begin{pweavecode}
hatx = FieldVector([1./3., 1./3.])
maxValue = max(f(e, hatx) for e in f.grid.elements)
\end{pweavecode}

In fact, since \pyth{function} is based on a globally defined function,
the following also works
\renewcommand{\pweavecaption}%
{Using a global evaluate method on grid function objects}
\renewcommand{\pweavelabel}{lstGF5}

\begin{pweavecode}
maxValue = max(f(e.geometry.toGlobal(hatx)) for e in f.grid.elements)
\end{pweavecode}

Note, that the evaluation in a global coordinate will not be available
for the \pyth{hat0} object defined in Listing~\ref{lstGF1}.

There is also a decorator \pyth{dune.grid.GridFunction} available to
decorate a class providing either a method \pyth{__call__(self, x)} or a
method \pyth{__call__(self, element, hatx)}.
In general, however, it will be more efficient to implement the full grid
function interface to cache element information upon \pyth{bind}.

For convenience, grid functions can be easily plotted using the \matplotlib:
\renewcommand{\pweavecaption}{Plotting grid functions}
\renewcommand{\pweavelabel}{lstGF6}

\begin{pweavecode}
f.plot()
hat0.plot()
\end{pweavecode}
\begin{figure}[htpb]
\center
\includegraphics[width= 0.45\textwidth]{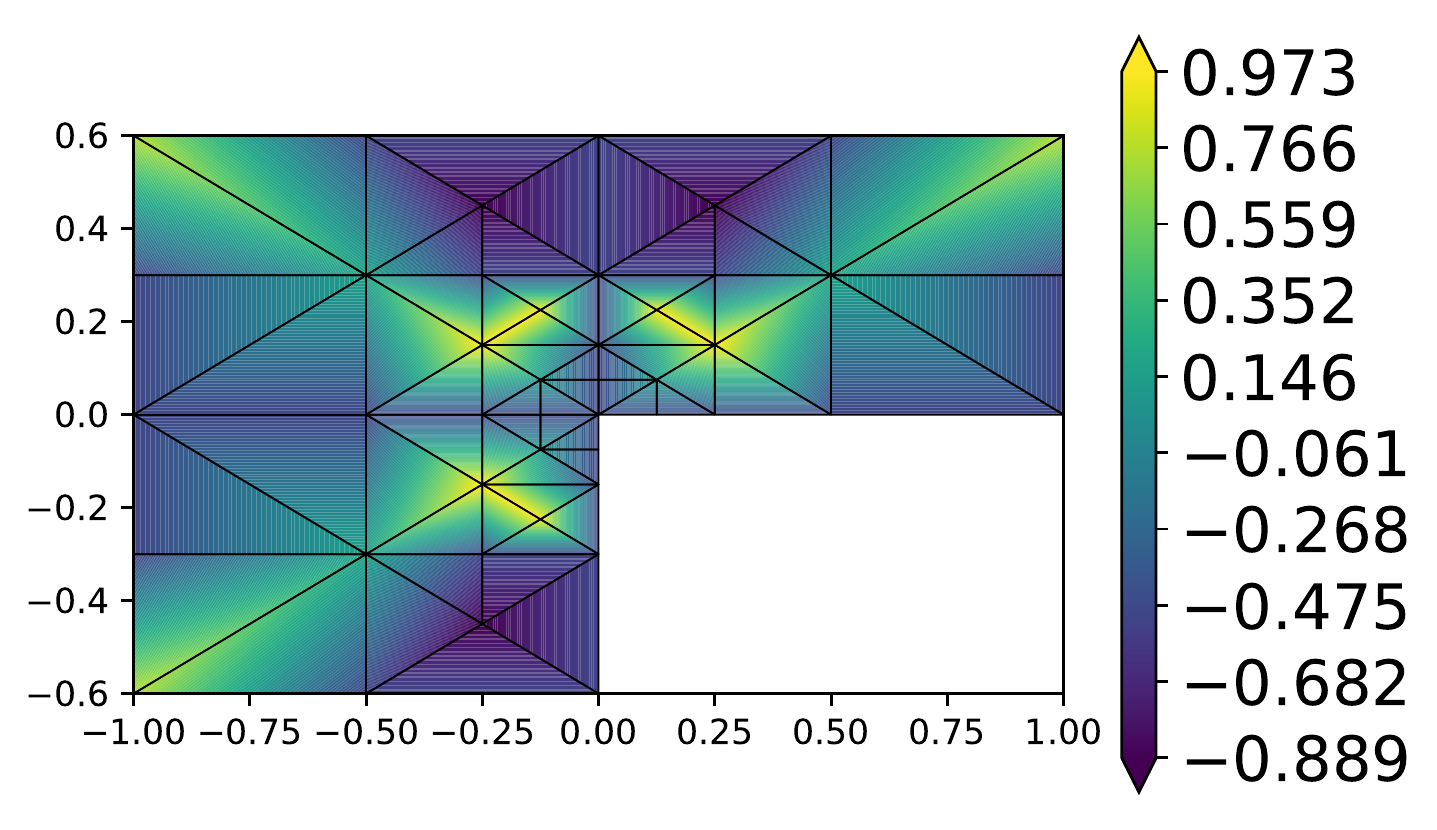}
\includegraphics[width= 0.45\textwidth]{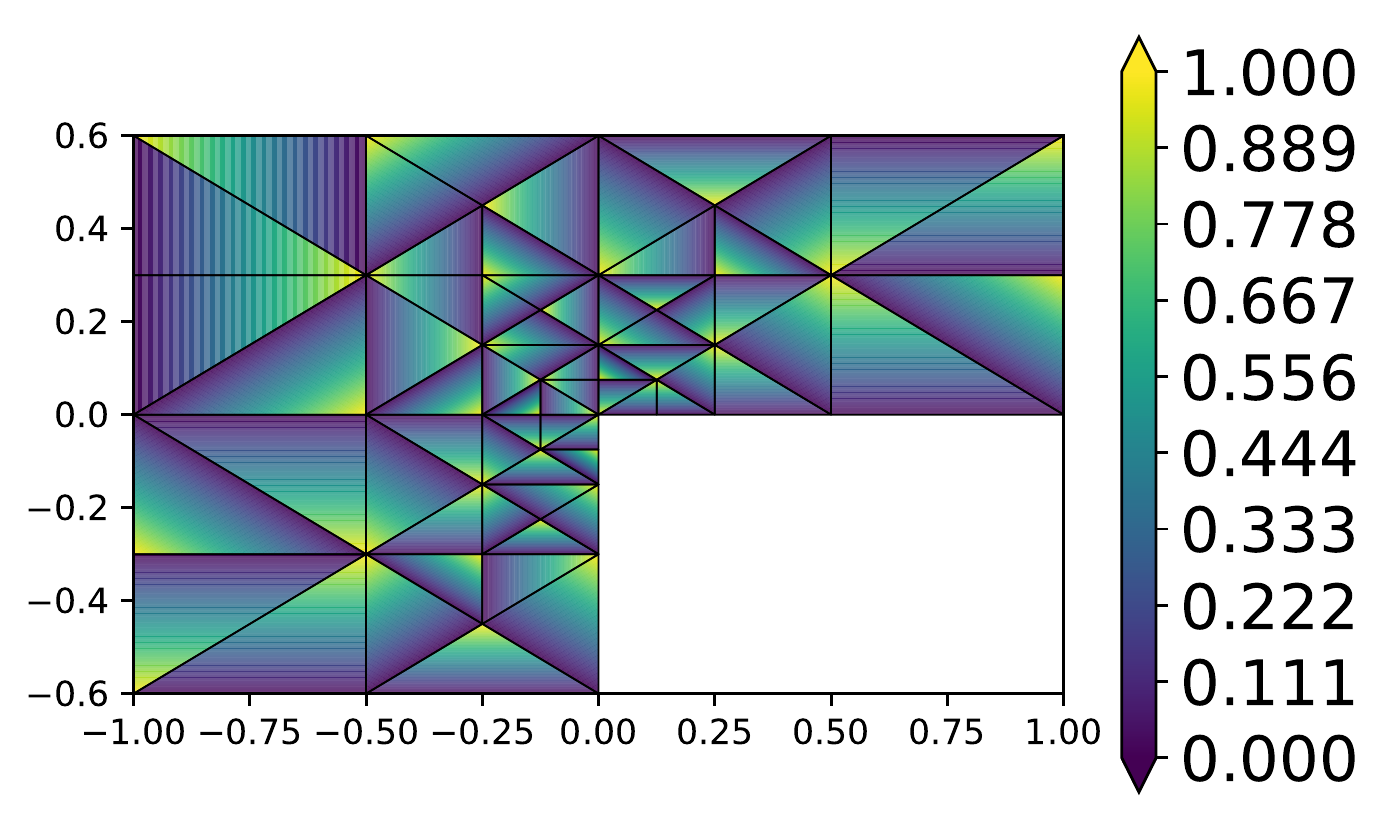}
\caption{Plot of \pyth{f} (left) and \pyth{hat0} (right)}
\label{fig:GlobalGridFunctions4}
\end{figure}



\subsection{Attaching Data to the Grid}
\label{sec:attachingdata}

In \dune grid structure and user data are cleanly separated, i.e.,
user data is stored in separate containers.
The grid provides index maps assigning a unique index to each entity,
which can be used to address containers of user data.

Each grid has an \pyth{indexSet} property, which assigns an index to each
entity that is unique within its geometry type.
For each geometry type, the index range is $[0, N)$, where $N$ is the number
of entities of this type within the grid.
Note that \dune does not prescribe any correlation between the iteration order
and the order of the indices. Recall, for example, the code snippet from
Listing~\ref{lstCreateFromGrid}, where the \pyth{indexSet} property was used to
define the elements of a triangulation. The function call
\pyth{aluView.indexSet.subIndices(e, 2)} for a given element \pyth{e} returns
the list of indices for all the corners of $e$ (the second argument being the codimension of the
vertices in a 2d grid; the indices for all the edges for example are obtained by calling
\pyth{subIndices(e,1)}).

To simplify attaching data to entities of different geometry type
(e.g., different codimension), the \dune[Grid] module provides the
\cpp{MultipleCodimMultipleGeomTypeMapper}.
It combines the ranges in an index set such that each entity in the grid is
assigned a fixed number $n$ of array indices, where $n$ depends only on the
geometry type.

In Python, the grid provides a convenience method \pyth{mapper} to construct
such a mapper.
The mapping from geometry type to the number of requested array indices can be
passed either as a function, as a callable, as a \pyth{dict}, or as a list or
tuple.
In the latter case, the number is assigned by codimension only.
For example, to assign 2 indices to each element and 3 indices to each vertex in
the grid, we construct the mapper as follows:
\setpweavecaption{Example mapper construction}{lstMapperConstruction}

\begin{pweavecode}
mapper = unitSquare.mapper([2, 0, 3])
\end{pweavecode}

The above approach is convenient if the same number of degrees of freedom
is attached to all subentities of a given codimension.
In many cases, however, the number depends on the geometry type of the entity.
For example, to attach 4 degrees of freedom to a quadrilateral but only 1 to a
triangle we use the following \pyth{dict}:
\setpweavecaption{Example mapper construction using a dictionary}{lstMapperConstructionDict}

\begin{pweavecode}
layout = {dune.geometry.quadrilateral: 4, dune.geometry.triangle: 1}
mapper = unitSquare.mapper(layout)
\end{pweavecode}

Note the abbreviations \pyth{dune.geometry.triangle} and
\pyth{dune.geometry.quadrilateral} which can be used for
\pyth{dune.geometry.simplex(2)} and
\pyth{dune.geometry.cube(2)}, respectively, as discussed in
Section~\ref{sec:CommonAndGeometry}.

Implementing a Lagrange type interpolation into a piecewise linear
finite element space can now be easily done:
\setpweavecaption{Lagrange interpolation}{lstLagrangeInterpolation}

\begin{pweavecode}
def interpolate(grid):
    mapper = grid.mapper({dune.geometry.vertex: 1})
    data = numpy.zeros(mapper.size)
    for v in grid.vertices:
        data[mapper.index(v)] = f(v.geometry.center)
    return mapper, data
\end{pweavecode}

For a triangular grid, implementing the linear interpolation elementwise is now
straightforward using the \pyth{gridFunction} decorator to obtain a
grid function:
\renewcommand{\pweavecaption}%
{Defining the local interpolation and the interpolation error}
\renewcommand{\pweavelabel}{lstInterpolationAndError}

\begin{pweavecode}
mapper, data = interpolate(aluView)
@dune.grid.gridFunction(aluView)
def p12dEvaluate(e, x):
    bary = 1-x[0]-x[1], x[0], x[1]
    idx = mapper.subIndices(e, 2)
    return sum(b * data[i] for b, i in zip(bary, idx))
\end{pweavecode}

In the above code we have used the \pyth{subIndices} method, which takes an
element \pyth{e} of the grid and a codimension \pyth{c} and returns the indices of all
degrees of freedom attached to subentities of \pyth{e} of the given
codimension (here, 2 for the vertices). Previously, we used the
\pyth{index} method providing an entity of the grid (a vertex) which
returns the indices of the degrees of freedom attached to that entity.

Finally, the mapper is callable with an element \pyth{e} to obtain
the indices of \emph{all} degrees of freedoms attached to that
element, i.e., including all subentities. The order of the returned indices
is in decreasing order of codimension, i.e., starting with the indices for
vertices.
Within a given codimension the indices are ordered according
to the order of the subentities in the reference element of \pyth{e}.

The maximum error at element barycenters can now be quite easily computed:
\renewcommand{\pweavecaption}%
{Maximum error of Lagrange interpolation using local coordinate object}
\renewcommand{\pweavelabel}{lstMaxErrorLagrangeInterpolation1}

\begin{pweavecode}
@dune.grid.gridFunction(aluView)
def error(e, x):
    return abs(p12dEvaluate(e, x)-f(e, x))
hatx = FieldVector([1./3., 1./3.])
print(max(error(e, hatx) for e in aluView.elements))
\end{pweavecode}
\begin{pweaveout}
1.6026566867981322
\end{pweaveout}

\subsection{Output}
\label{sec:output}

Let us first visualize the result using \matplotlib. We have already
used the \pyth{plot} method on the grid class. The \pyth{gridFunction}
decorated also adds a plot method that can be used to plot a piecewise
linear interpolation of the data:
\renewcommand{\pweavecaption}%
{Plotting the Lagrange interpolation}
\renewcommand{\pweavelabel}{lstPlotLagrangeInterpolation}

\begin{pweavecode}
p12dEvaluate.plot(figsize=(9,9), gridLines=None)
p12dEvaluate.plot(figsize=(9,9), gridLines='black',
             xlim=[0,0.4], ylim=[0,0.4])
f.plot(level=2, figsize=(9,9), gridLines=None)
\end{pweavecode}
\begin{figure}[htpb]
\center
\includegraphics[width= 0.32\textwidth]{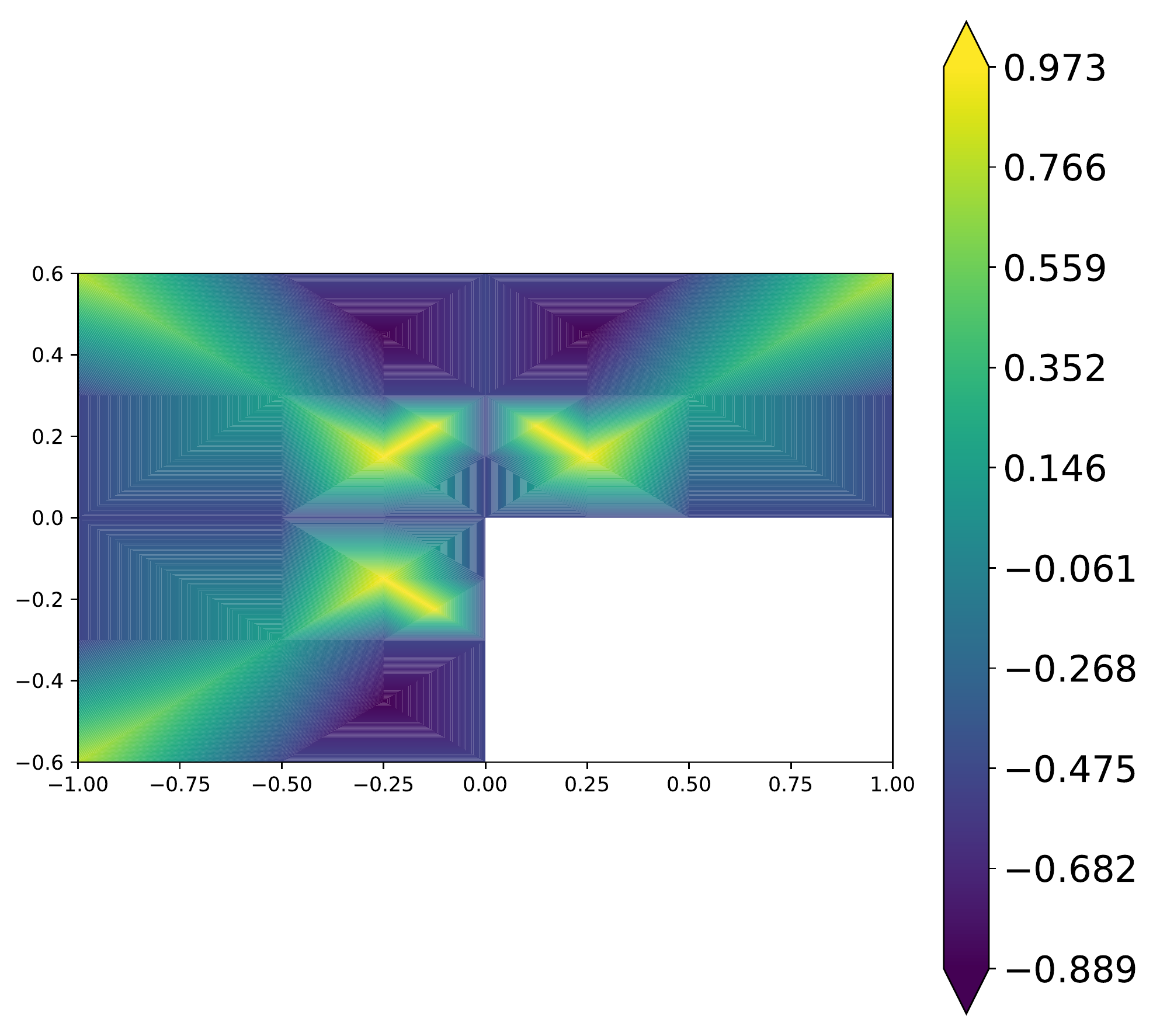}
\includegraphics[width= 0.32\textwidth]{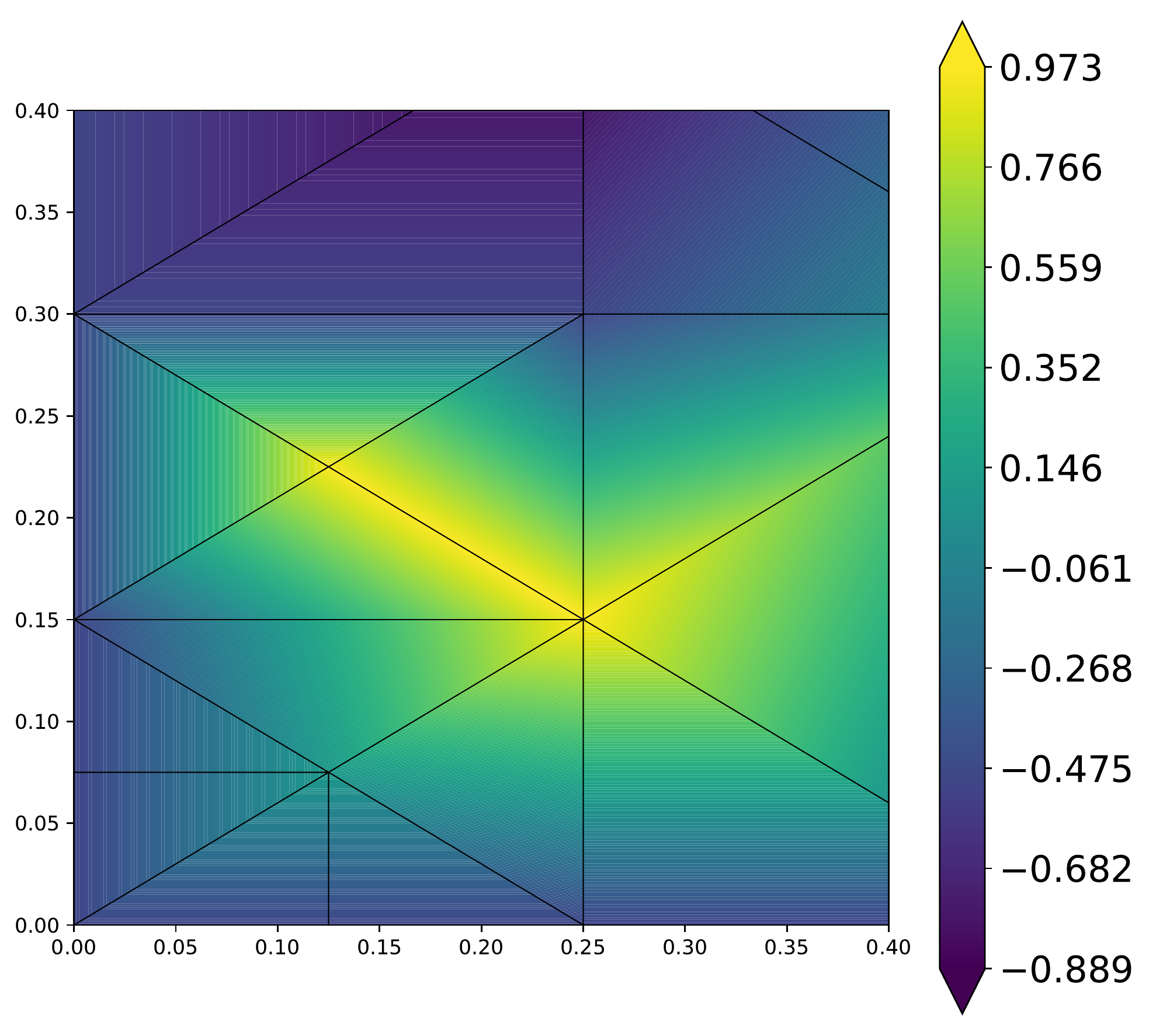}
\includegraphics[width= 0.32\textwidth]{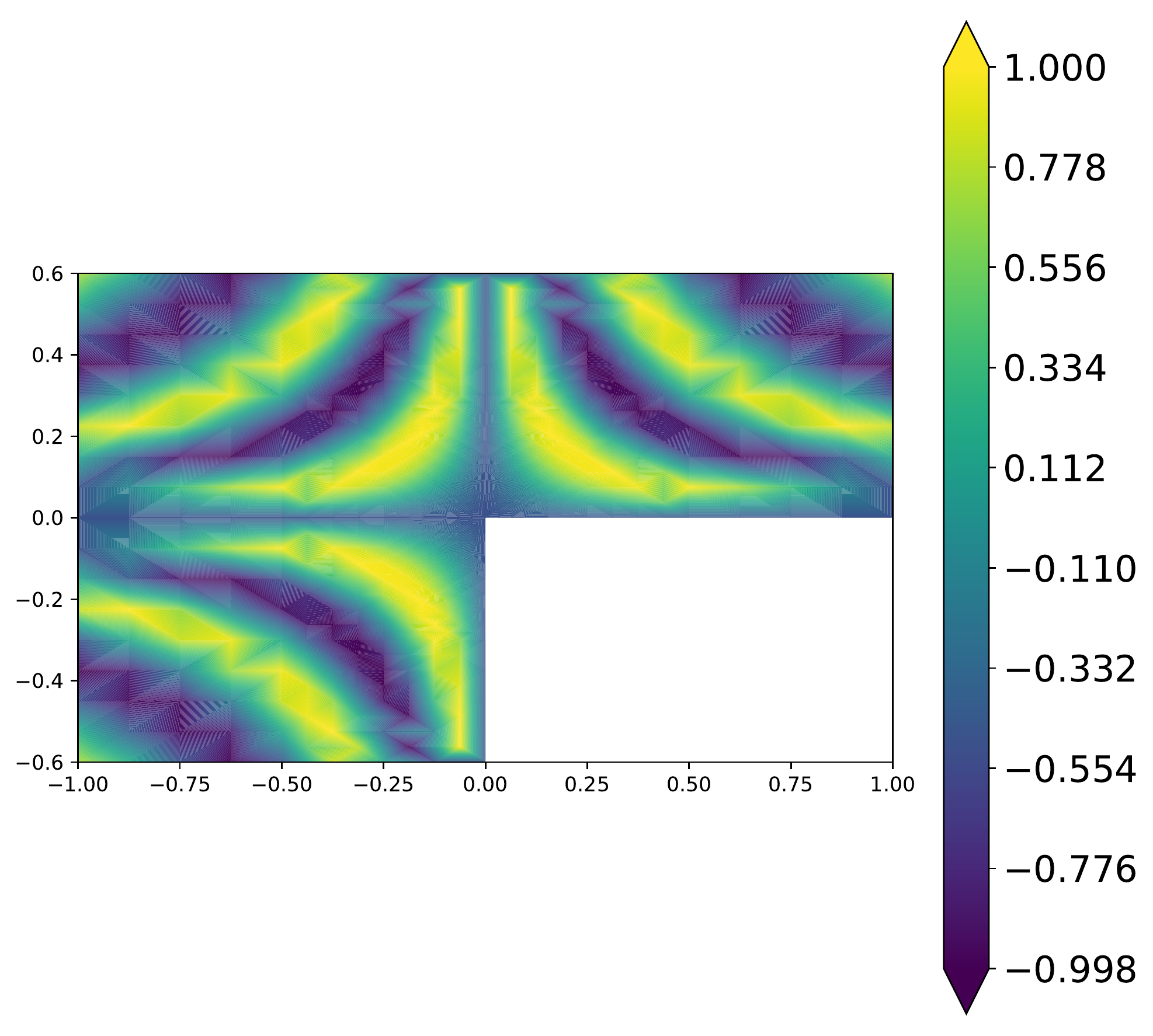}
\caption{Lagrange interpolation on the locally adapted grid (full domain on left and on $[0,0.4]^2$ on right)}
\label{fig:LagrangeInterpolationFig}
\end{figure}

In the final line we plot the actual function we are interpolating. We used
an additional argument \pyth{level} to get a more accurate representation.
If we had set \pyth{level=0} we would have reproduced the picture of the
linear interpolation, i.e., the figure on the left;
by using \pyth{level>0} the result is shown on a
refined grid. Note that both the locally defined function
\pyth{p12dEvaluate} and the globally defined function are handled in the
same way.

The capabilities of \matplotlib are limited to two-dimensional grids at most.
Moreover at the time of writing, only the grid's wireframe and piecewise linear
functions can be plotted (albeit also on a refined grid) using the provided
methods.
For more complex data analysis, e.g., in three space dimensions, external
programs like \paraview are more flexible.
Thus, \dune[Grid] provides output of grid and data using the \vtk file format.
The following code snippet produces a \vtk file with the piecewise linear
interpolation and the actual smooth function we interpolated:
\renewcommand{\pweavecaption}%
{\vtk output}
\renewcommand{\pweavelabel}{lstVTKOutput}

\begin{pweavecode}
pd = {"exact": f, "discrete": p12dEvaluate, "error": error}
aluView.writeVTK("interpolation", pointdata=pd)
\end{pweavecode}

As with \matplotlib, the error would be zero and we can use subsampling to see
the difference between the exact function and the linear interpolation in more
detail:
\renewcommand{\pweavecaption}%
{Subsampling \vtk output}
\renewcommand{\pweavelabel}{lstSubsamplingVTKOutput}

\begin{pweavecode}
aluView.writeVTK("interpolation_subsampled", subsampling=2, pointdata=pd)
\end{pweavecode}

\begin{figure}[htpb]
\center
\includegraphics[width=0.32\textwidth]{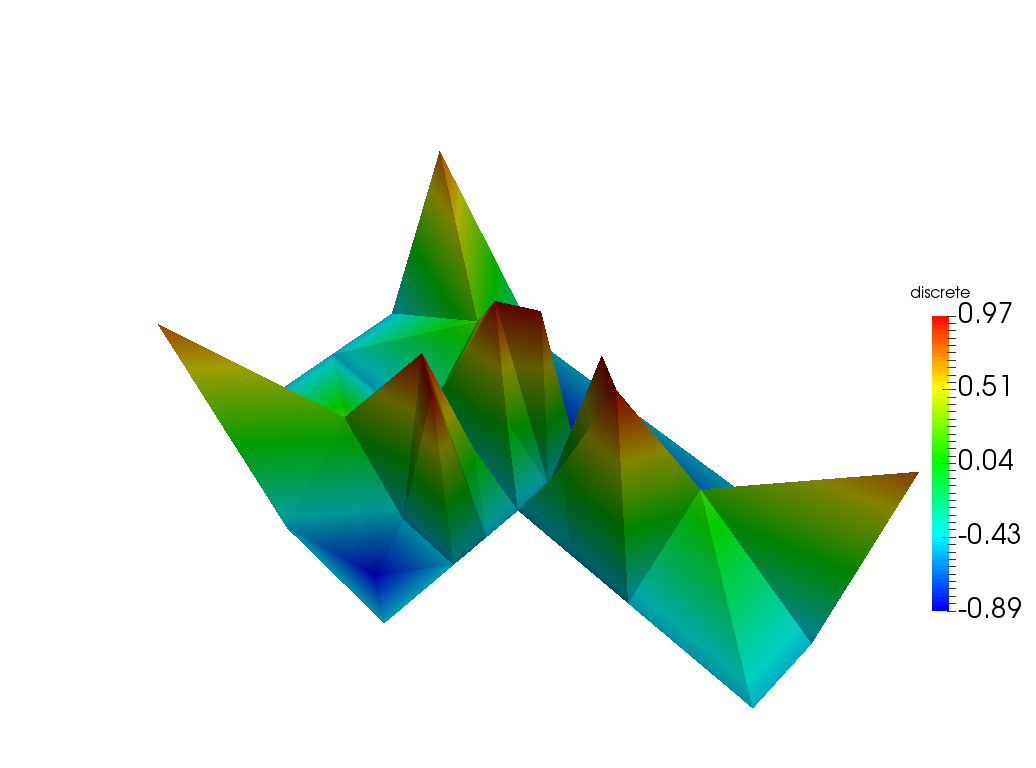}
\includegraphics[width=0.32\textwidth]{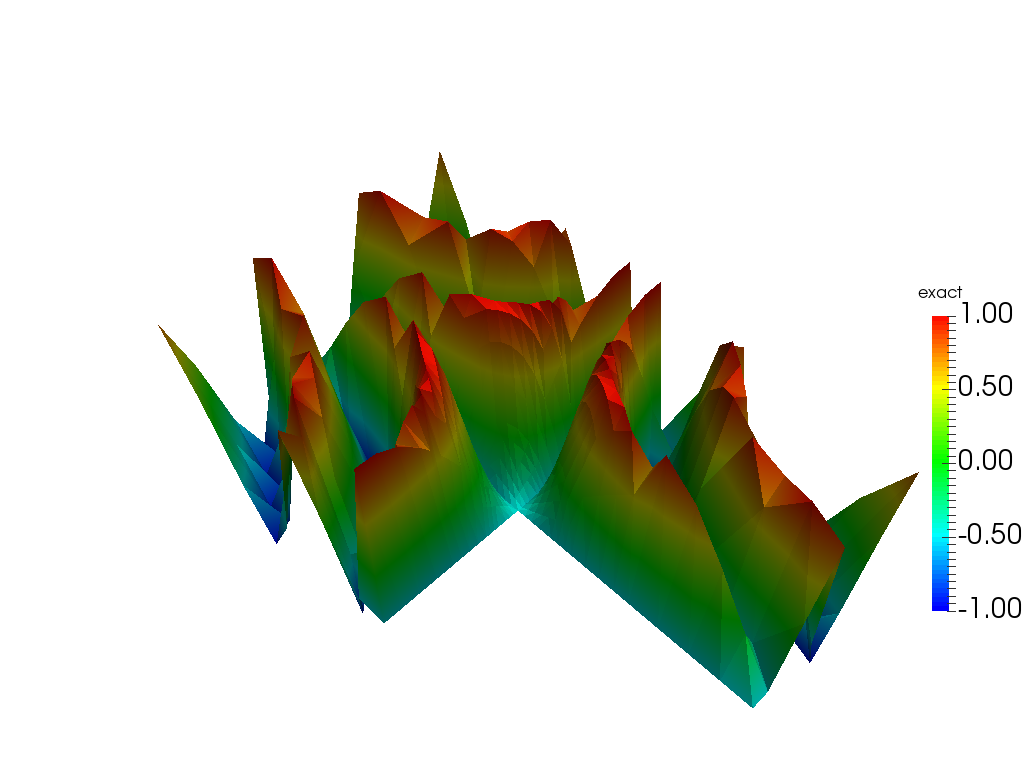}
\includegraphics[width=0.32\textwidth]{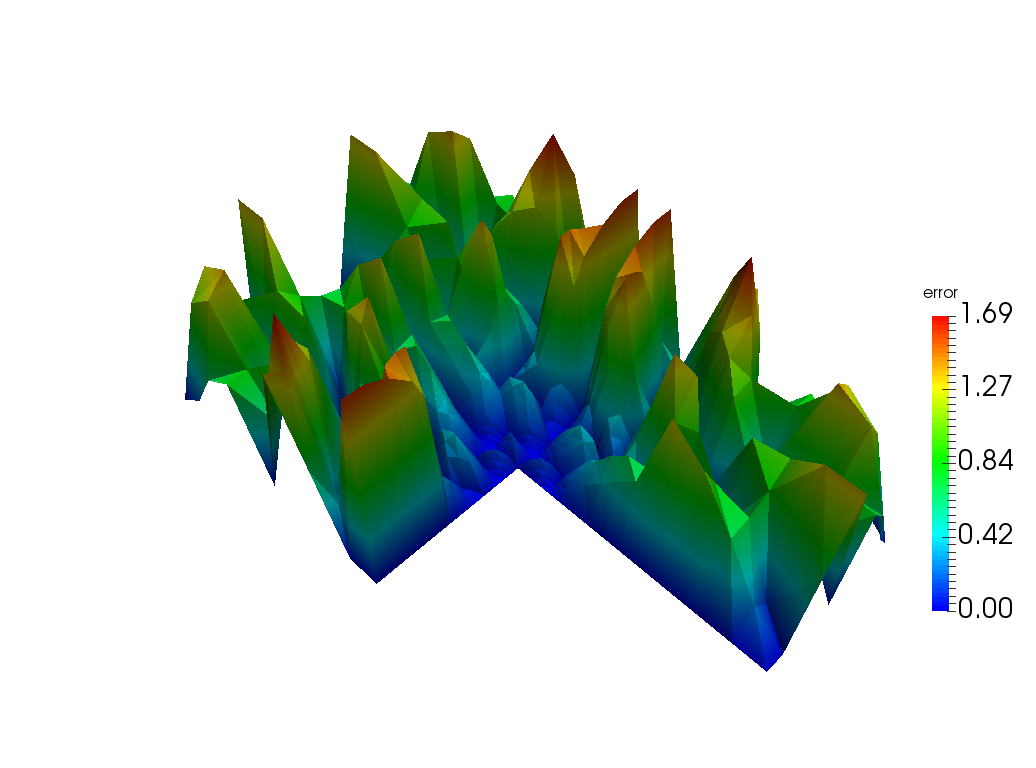}
\caption{Interpolation, exact function, and error using the vtk output}
\label{fig:interpolSubSampled}
\end{figure}

It is also directly possible to extract a representation of the
tessellation using \numpy structures using the \pyth{triangulation}
method, which returns a \pyth{matplotlib.tri.triangulation.Triangulation}
object.
A \numpy array containing the values of a grid function, e.g., the Lagrange
interpolation defined in Listing~\ref{lstLagrangeInterpolation},
at the nodes of the (subsampled) grid can be obtained using the method
\pyth{pointData} on the grid function.
The following example shows how this can be used to plot grid functions using
\mayavi \cite{mayavi}:
\renewcommand{\pweavecaption}%
{Converting tessellation and interpolation to \numpy}
\renewcommand{\pweavelabel}{lstToNumpy}

\begin{pweavecode}
level = 3
triangulation = f.grid.triangulation(level)
z = f.pointData(level)[:,0]
try:
    from mayavi import mlab
    from mayavi.tools.notebook import display
    mlab.init_notebook()
    mlab.figure(bgcolor = (1,1,1))
    s = mlab.triangular_mesh(triangulation.x, triangulation.y, z*0.5,
                             triangulation.triangles)
    display(s)
    mlab.savefig("mayavi.png", size=(400,300))
    mlab.close(all=True)
except ImportError:
    pass
\end{pweavecode}

\begin{figure}[htpb]
\center
\includegraphics[width=0.6\textwidth]{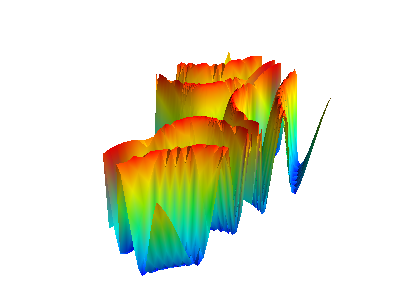}
\caption{Visualization using mayavi}
\label{fig:mayavi}
\end{figure}

\subsection{Parallelization}
\label{sec:parallelization}

The \dune framework provides support for distributed memory parallelization
based on MPI, though it is not required for serial computations.
The framework for basic communication are provided by \dune[Common].
For example, the \pyth{CollectiveCommunication} provides an interface to the
global communication patterns, e.g., global reduction operations.

In a parallel \dune grid, the elements are partitioned such that each element
is owned by exactly one process, referred to as the interior partition.
To support numerical algorithms, further elements might be known to each
process as overlap or ghost elements.

For global reduction opererations, the grid provides a
\pyth{CollectiveCommunication} object including all processes the grid is known
to through the property \pyth{comm}.
To perform data exchange between between processes sharing entities of any
codimension within a partition, the grid provides a method \pyth{communicate}
taking a structure describing how the data is to be gathered/scattered between
the neighboring processes.

Although this low level interface can be used through \dune[Python], we will
describe a simpler convenience approach provided in this module focusing on data
stored in \numpy arrays using the mapper described in
Section~\ref{sec:attachingdata}.

Before we describe the method for communication we need to give a brief
description of \dune's concept of grid partitions. Each process stores
part of the grid and assigns to each entity in the grid a unique partition
type: \cpp{interior}, \cpp{border}, \cpp{ghost}, \cpp{overlap}, and
\cpp{front}.
As described above, \cpp{interior} refers to all entities owned by the
process while \cpp{ghost} and \cpp{overlap} entities are copies of entities
owned by another process. In distinction, \cpp{ghost} entities provide a
reduced set of information compared to \cpp{overlap} entities.

Entities of higher codimension might be subentities of both, \cpp{interior} and
non-\cpp{interior} elements. These are assigned the partition type
\cpp{border}.
In contract, \cpp{front} entities are subentities of both, \cpp{overlap} and
\cpp{ghost} entities, but not \cpp{interior} elements.
Full details will not be required for the followingm, but for a in-depth
description of the concept we refer to \cite{dunepaperI:08}.

Communication is now
performed between entities on the sending and receiving processor of
a given partition type. To this end \dune provides the most commonly used
combinations of partition types:
\begin{itemize}
  \item
    \cpp{interiorPartition} for \cpp{interior} entities only,
  \item
    \cpp{interiorBorderPartition} for all \cpp{interior} and \cpp{border} entities,
  \item
    \cpp{overlapPartition} for \cpp{interior}, \cpp{border}, and \cpp{overlap} entities,
  \item
    \cpp{overlapFrontPartition} for all but \cpp{ghost} entities,
  \item
    \cpp{allPartition} for all entities known the the process.
\end{itemize}
It is now possible to iterate over these subsets in \dune[Python] as follows.
\renewcommand{\pweavecaption}%
{Iteration over the elements in the \pyth{interiorBorderPartition} of a given grid view}
\renewcommand{\pweavelabel}{lstPartitionIterator}

\begin{pweavecode}
for entity in aluView.interiorBorderPartition.elements:
    pass
\end{pweavecode}

Now, data is exchanged by gathering the data attached to entities of a
\pyth{fromPartition} and scattering it to a \pyth{toPartition}.
The corresponding communication method is provided by the `mapper` class and
takes any number of \numpy arrays storing data using the mapper instance:
\renewcommand{\pweavecaption}{Communicate method on the mapper class}
\renewcommand{\pweavelabel}{lstMapperComm1}
\begin{pweavecode}
mapper.communicate(toPartition, fromPartition, operation, data1, ..., dataN)
\end{pweavecode}
Here \pyth{operation} is a function reducing a local and a remote value to one,
called upon receiving data, specifying how to combine the local value with a
value receive from another process.
Two reduction operations are frequently used
\emph{set} (\pyth{lambda local,remote: remote})
and \emph{add} (\pyth{lambda local,remote: local+remote}).
For these two we provide a simplified (and optimized) notation, e.g.,
\renewcommand{\pweavecaption}%
{Communicate method on the mapper class from \pyth{interior+border} to
\pyth{all} elements using predefined operation \pyth{set}}
\renewcommand{\pweavelabel}{lstMapperComm2}

\begin{pweavecode}
mapper.communicate(aluView.interiorBorderPartition, aluView.allPartition, \
                   dune.grid.CommOp.set, data)
\end{pweavecode}

The arguments \pyth{fromPartition} and \pyth{toPartition} can either be one of
the predefined partitions. Notice, however, that not all combinations are legal,
e.g., you can neither send or receive on the \pyth{interiorPartition} nor can
you combine the \pyth{interiorBorderPartition} with the \pyth{overlapPartition}
or the \pyth{overlapFrontPartition}.
The following code gives a full working example
where each vertex $v$ is assigned the minimum rank of all processors holding
copy of $v$:
\inputpython{communicate.py}{Communication example}{lstMapperComm3}
\begin{figure}[htpb]
\center
\includegraphics[width=0.8\textwidth]{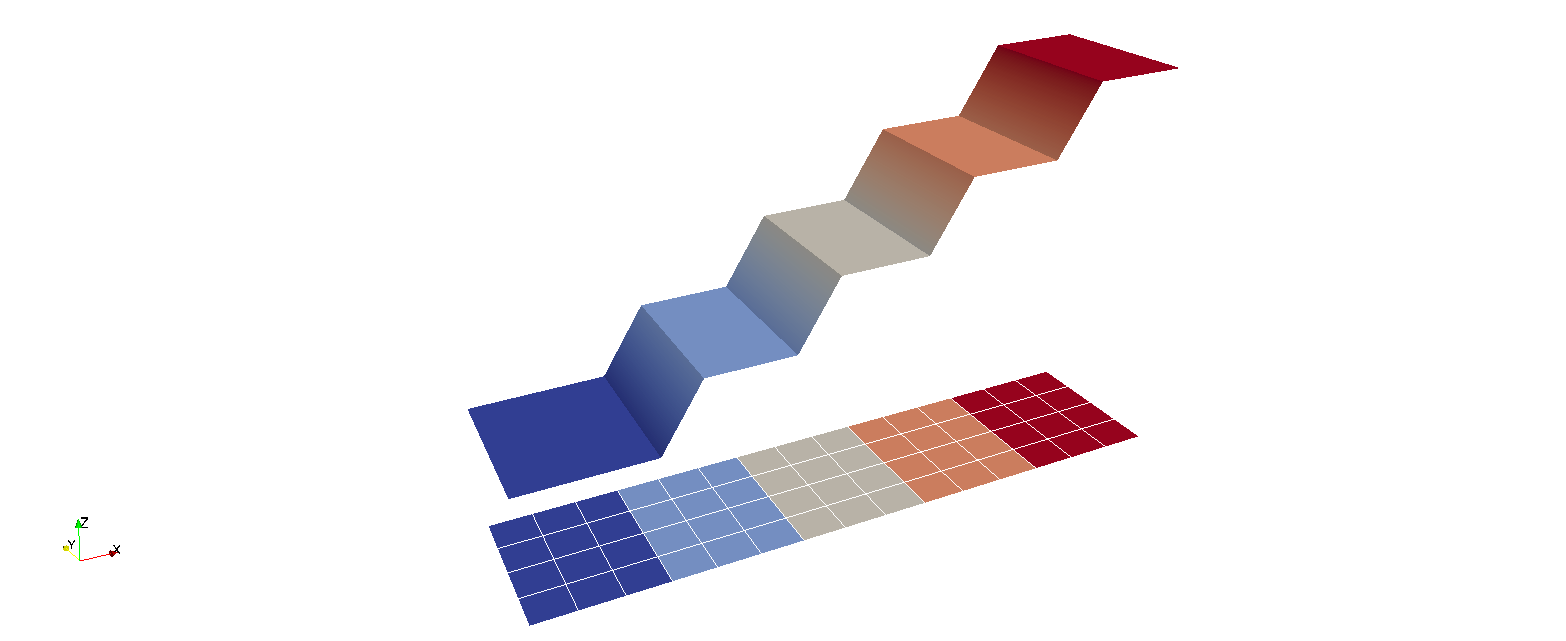}
\caption{Communication example where each vertex $v$ is assigned the minimum rank
         holding a copy of $v$ using five processors.}
\label{fig:communicate}
\end{figure}

\section{Vectorization support}
\label{sec:vectorization}

As pointed out above one of the aims of the Python interface provided in
this module, is to make the transition as straightforward as possible both
for users of the C++ interface to get started using \dune[Python] and for
experienced Python developers to transfer their Python algorithms to C++.
The discussion of the Python interface so far focused on this aspect
and on the minor changes we decided to make when exporting the C++
interface to Python. Mostly the changes involve
using properties instead of functions and in some cases providing a Python
magic method (e.g., \pyth{__str__} or \pyth{__len__}); in cases where there
is an analogous method in the C++ interface (e.g., \cpp{size}) both the
magic method and the C++ method are provided to allow both for a Pythonic
experience and a close match of the C++ interface for experienced \dune
users.

An issue with mimicking the C++ interface is that this can lead to low
performance code. We are not aiming at a very high level of
performance of the pure Python code (to that end the \pyth{algorithm}
module provides JIT functionality as described in Section \ref{sec:JIT}).
Nevertheless, we add two extensions to the Python interface which result in
significant improvements in performance. We have already been using the first
approach throughout the previous section: on the C++ side there are many
methods taking an index ranging over a small (often fixed) range. For
example the method \cpp{corner(i)} on the \cpp{Geometry} interface which
returns the coordinates of the ith corner. Obtaining all corners then
requires a loop over \cpp{i} and the range for this index is returned by
the matching method \cpp{corners}. Implementing a loop in this form on the
Python side leads to a significant overhead due to the frequent calls from
Python into C++. On the C++ side we can rely on compiler optimization (e.g.,
inlining) to produce optimal code but this will not be the case on the
Python side. For this reason the \pyth{corners} method on the Python side
directly returns a tuple containing the coordinates of all corners,
so the correct loop reduces to \pyth{for c in geometry.corners} and only
requires a single call from Python to C++.

The second approach used to increase performance is to apply some
of the C++ methods not to a single
input but to a vector of inputs. Again by simply reducing the number of calls
between Python and C++ we can greatly improve the performance
of the Python code without causing an unacceptable deviation from the C++
\dune interface. In Python this is often referred to as vectorization and
is heavily used for example in code using \numpy.
So far, we have extended the methods on the \cpp{geometry}
to handle multiple coordinates stored in \numpy arrays as input and
provided vectorization support for the evaluation of grid functions.

Since many central parts of numerical methods require the approximation of
integrals over the elements in the grid, we focus on numerical quadrature
to explore the vectorization capabilities available in \dune[Python] at
the time of writing. The module \dune[Geometry] provides a number of
quadrature rules which have been exported to \dune[Python] and which we
will use in the following to demonstrate the use of vectorization.

Using the \dune[Python] method described so far, the following code shows
how to compute the $L^2$ error of the Lagrange interpolation $u_h$ of
$u=\cos(\frac{2\pi}{0.3+xy})$ from
Code Listing~\ref{lstLagrangeInterpolation}.
We need to use \numpy to define the function and, for the sake of completeness,
we also recall the definition of the interpolation and the error, although they
don't require any changes.
To get more meaningful results we also reduce the grid spacing and,
therefore, need to recompute the degrees of freedom for the interpolant:
\renewcommand{\pweavecaption}%
{Computing the Lagrange interpolation and the error}
\renewcommand{\pweavelabel}{lstPyIntegral1}

\begin{pweavecode}
@dune.grid.gridFunction(aluView)
def function(x):
    return numpy.cos(2.*numpy.pi/(0.3+abs(x[0]*x[1])))

aluView.hierarchicalGrid.globalRefine(4)
mapper, data = interpolate(aluView)

@dune.grid.gridFunction(aluView)
def p12dEvaluate(element,x):
    indices = mapper(element)
    bary = 1-x[0]-x[1], x[0], x[1]
    return sum( bary[i] * data[indices[i]] for i in range(3) )

@dune.grid.gridFunction(aluView)
def error(element,x):
    return numpy.abs(p12dEvaluate(element,x)-function(element,x))
\end{pweavecode}

Now we compute
$$ \sum_{E\in{\cal T }_h}
          \sum_{(\hat{\omega},\hat{x})\in Q}
               \hat{\omega} |{\rm det DF_E(\hat{x})}|
               |u(F_E(\hat{x}))-u_h(F_E(\hat{x})|^2
   \approx \int_\Omega |u-u_h|^2 $$
where ${\cal T}_h$ denotes the grid and $F_E$ is the mapping from the reference
element $\hat E$ of the element $E \in {\cal T}_h$ to $E$, modeled by the
\pyth{Geometry} class.
To compute the integral over $E$ we use a quadrature $Q=\{(\hat{\omega},\hat{x})\}$ for the
reference element:
\renewcommand{\pweavecaption}%
{Computing the $L^2$ error of the Lagrange interpolation}
\renewcommand{\pweavelabel}{lstPyIntegral2}

\begin{pweavecode}
start = time.time()
l2norm2 = 0
for e in aluView.elements:
    geo = e.geometry
    for p in dune.geometry.quadratureRule(e.type, 5):
        hatx, hatw = p.position, p.weight
        weight = hatw * geo.integrationElement(hatx)
        l2norm2 += error(e,hatx)**2 * weight
print("L2 error of Lagrange interpolation:",math.sqrt(l2norm2))
print("time used:", round(time.time()-start,2))
\end{pweavecode}
\begin{pweaveout}
L2 error of Lagrange interpolation: 0.01933669460592608
time used: 3.07
\end{pweaveout}

This code is practically identical to its C++ counterpart but the large
number of calls of C++ methods have a significant impact on the performance.
A significant performance increase can be obtained by simply evaluating the
\pyth{error} function on all quadrature points simultaneously:
\renewcommand{\pweavecaption}%
{Computing the $L^2$ error of the Lagrange interpolation using vectorization}
\renewcommand{\pweavelabel}{lstPyIntegralVectorize1}

\begin{pweavecode}
start = time.time()
l2norm2 = 0
for e in aluView.elements:
    hatxs, hatws = dune.geometry.quadratureRule(e.type, 5).get()
    weights = hatws * e.geometry.integrationElement(hatxs)
    l2norm2 += numpy.sum(error(e, hatxs)**2 * weights, axis=-1)
print("L2 error of Lagrange interpolation:",math.sqrt(l2norm2))
print("time used:", round(time.time()-start,2))
\end{pweavecode}
\begin{pweaveout}
L2 error of Lagrange interpolation: 0.01933669460592627
time used: 0.99
\end{pweaveout}

As pointed out computing integrals is a central part of many numerical
schemes, e.g., computing load vectors and stiffness matrices in addition to
computing errors. Consequently, we have added a function to compute an
integral over a given element:
\renewcommand{\pweavecaption}{One line computation of the error integral}
\renewcommand{\pweavelabel}{lstPyIntegralOneLine}

\begin{pweavecode}
from dune.geometry import integrate
start = time.time()
rules = dune.geometry.quadratureRules(5)
l2norm2 = sum(integrate(rules, e, lambda e, x: error(e, x)**2)
              for e in aluView.elements)
print("One line approach:", math.sqrt(l2norm2))
print("time used:", round(time.time()-start,2))
\end{pweavecode}
\begin{pweaveout}
One line approach: 0.01933669460592627
time used: 0.96
\end{pweaveout}

\dune[Python] also provide bindings for the Python module
\quadpy \cite{QuadPy} which provides a wide range of quadrature rules:
\renewcommand{\pweavecaption}%
{Computing the $L^2$ error of the Lagrange interpolation}
\renewcommand{\pweavelabel}{lstQuadPyIntegral1}

\begin{pweavecode}
try:
    import dune.geometry.quadpy as quadpy
    start = time.time()
    qrules = quadpy.rules({dune.geometry.triangle: (5, "XiaoGimbutas")})
    l2norm2 = sum(integrate(qrules, e, lambda e, x: error(e, x)**2)
                  for e in aluView.elements)
    print("Using QuadPy:", math.sqrt(l2norm2))
    print("time used:", round(time.time()-start,2))
except ImportError:
    pass
\end{pweavecode}
\begin{pweaveout}
Using QuadPy: 0.01933669460592627
time used: 3.84
\end{pweaveout}

A further increase in performance can be obtained by using the
\pyth{dune.generator.algorithm} module introduced in Section~\ref{sec:JIT}.
We show here two approaches, the first using standard \dune
methods which require quite a number of calls to the Python function
\pyth{error}, the other using the vectorization features also available on
the C++ side when using \dune[Python]:
\renewcommand{\pweavecaption}%
{Computing the $L^2$ error of the Lagrange interpolation using JIT
compilation}
\renewcommand{\pweavelabel}{lstPyIntegral333}

\begin{pweavecode}
from dune.generator import algorithm

start = time.time()
l2norm2 = algorithm.run('l2norm2a', 'l2norm2.hh', aluView, error)
print("A: L2 error of Lagrange interpolation:",math.sqrt(l2norm2))
print("time used:", round(time.time()-start,2))

start = time.time()
l2norm2 = algorithm.run('l2norm2b', 'l2norm2.hh', aluView, rules, error)
print("B: L2 error of Lagrange interpolation:",math.sqrt(l2norm2))
print("time used:", round(time.time()-start,2))
\end{pweavecode}
\begin{pweaveout}
A: L2 error of Lagrange interpolation: 0.01933669460592608
time used: 2.27
B: L2 error of Lagrange interpolation: 0.01933669460592608
time used: 0.71
\end{pweaveout}

The C++ code is in the header file \file{l2norm.hh}
and contains the method
\begin{c++}{}{}
template< class GridView, class GF >
double l2norm2a ( const GridView &gridView, const GF& gf )
\end{c++}
performing the actual computation of the $L^2$ norm for a given grid function.
Note that this function can be used directly from within any \dune program based
based purely on the C++ interface.
The second version, \cpp{l2norm2b}, uses \numpy arrays to efficiently call
back into Python.


Finally, to get an impression on the overall performance, we include a
version completely written in C++.
While the implementation does use the \numpy vector for the vertex values, the
analytic function and the integration are implemented in pure C++, leaving no
flexibility to the Pyhton user.
Achieve this additional flexibility would require actual code generation, which is not
included in \dune[Python], but together with other more high level binding code is part
of the \dune[Fempy] module.

\renewcommand{\pweavecaption}%
{Completely compute the $L^2$ error of the Lagrange interpolation on the C++ side}
\renewcommand{\pweavelabel}{lstPyIntegral4}

\begin{pweavecode}
start = time.time()
headers = ['l2norm2.hh','fulll2norm2.hh']
l2norm2 = algorithm.run('l2norm2', headers, aluView, mapper, data)
print("L2 error of Lagrange interpolation:", math.sqrt(l2norm2))
print("time used:", round(time.time()-start,2))
\end{pweavecode}
\begin{pweaveout}
L2 error of Lagrange interpolation: 0.019336694605926106
time used: 0.27
\end{pweaveout}

In addition to the header file \file{l2norm2.hh}, this code uses the header
file \file{fulll2norm2.hh}.
Note that, despite the use of \numpy vectors, writing this code does not require
any understanding of \pybind.

\section{Examples}
\label{sec:examples}

In the following we present two standard numerical schemes:
a finite element scheme for solving linear elliptic problems
and a finite volume scheme for solving a linear transport equation.
The code follows the \dune[Grid-HowTo] examples, but uses \numpy for
storing vectors and \scipy for solving the resulting linear systems of
equations.


\subsection{Finite Element Scheme}

We start by investigating the efficiency of the vectorization approach using
the standard task of finite element assembly.
We use both load vector and stiffness matrix assembly on a conforming
triangulation using conforming second order Lagrangian finite elements:
\begin{align*}
  l_i    = \int_\Omega f\varphi_i~, \qquad
  A_{ij} = \int_\Omega \nabla\varphi_j\cdot\nabla\varphi_i~,
\end{align*}
where the $\varphi_i$ denote the basis functions.

Note that the C++ implementation for computing the load vector has one callback
to Python per element to obtain the values of the function $f$ at all quadrature
points. This results in a performance loss in the C++ code compared to a
pure C++ implementation; the vectorized Python code is about a factor of
$10$ slower. The pure Python version is about a further factor of $2$
slower and the complexity grows with the number of quadrature points while
the vectorized version hardly depends on the order of the quadrature.
In the case of the matrix assembly there is no callback required (we do not
study a varying diffusion model). Therefore, the difference between the C++
version and the vectorized implementation is significantly more pronounced.
The difference between the two Python versions is again about $2$ depending
on the quadrature order used.

\begin{figure}[htpb]
\center
\includegraphics[width=0.47\textwidth]{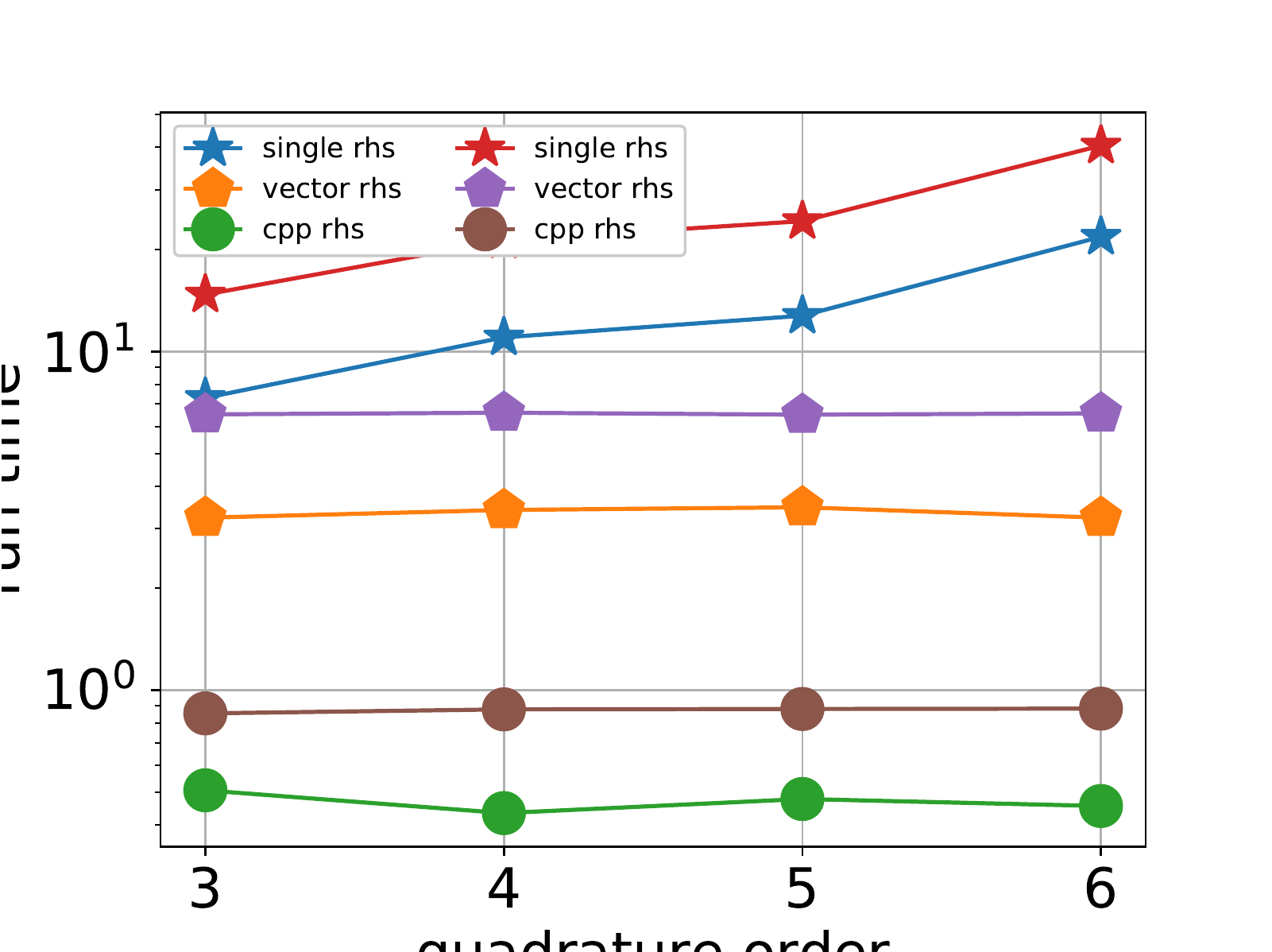}
\includegraphics[width=0.47\textwidth]{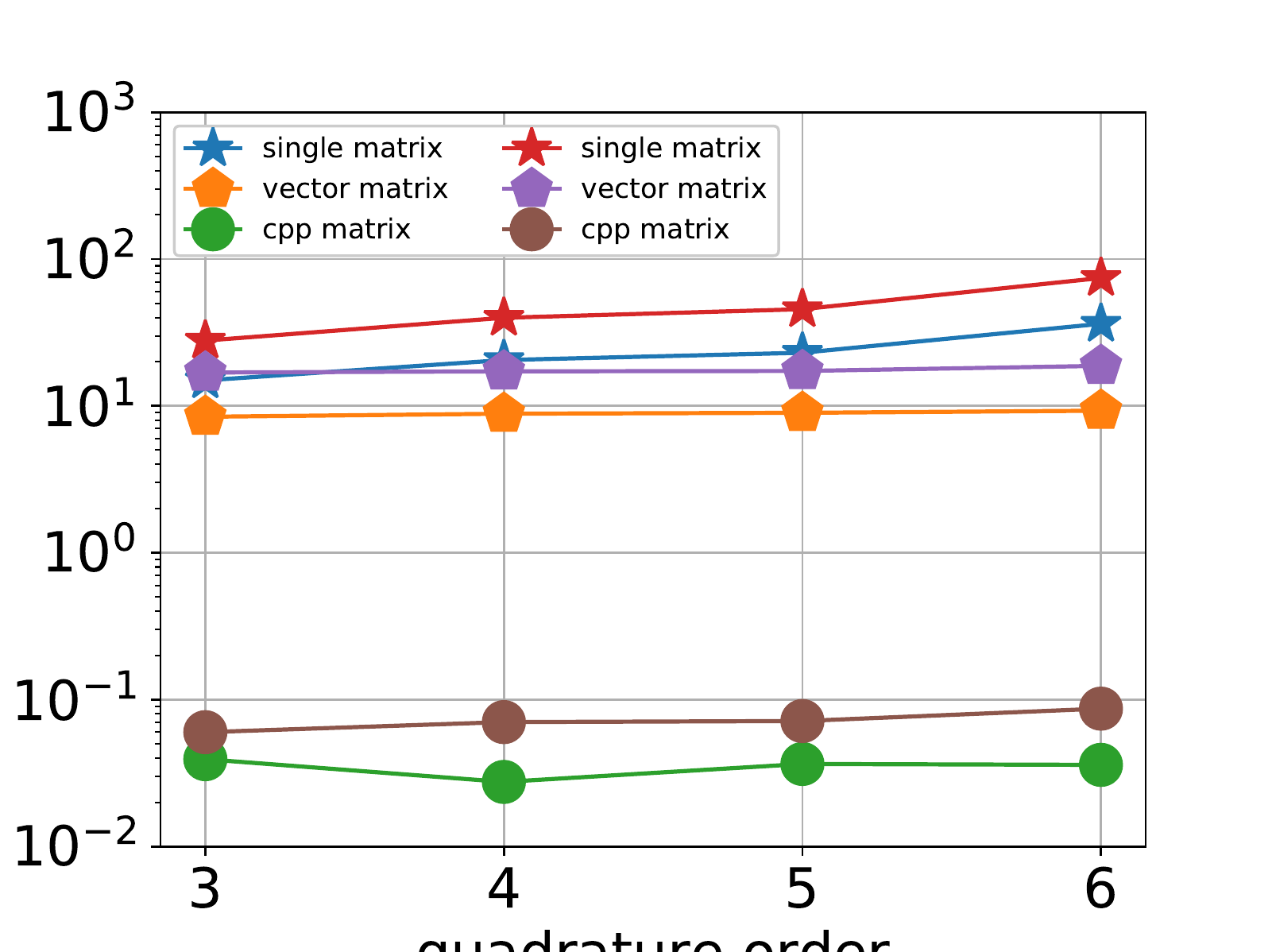}
\caption{Efficiency improvement by using vectorization for load vector
(left) and stiffness matrix (right) assembly for second order finite
elements. Results are shown for two different grid resultions.}
\label{fig:assemblycompcost}
\end{figure}


\subsection{Finite Volume Scheme}

First we investigate the computational cost for a finite volume scheme
for solving a simple linear transport problem
\begin{equation*}
  \partial_t u + \partial_x u + \partial_y u = 0,
\end{equation*}
with initial and inflow boundary data given by a function
$\bar{u}(t,x,y)$. The solution to this problem is approximated by a
piecewise constant function
$u^n_E\approx\frac{1}{|E|}\int_E u(n\tau,x)\;dx$ where $\tau>0$ is the time
step, $n>0$, and $E$ is an element of the grid.
An explicit scheme for solving this problem is given by
\begin{align*}
   u^0_E &= \bar{u}(0,\omega_T)  \\
   u^{n+1}_E &= u^n_E -
     \frac{\tau}{|E|}\sum_{\dim(E' \cap E) = 1} g(u^n_E,u^n_{E'},n_E)
\end{align*}
using an upwind flux $g$.

We first show results for a simulation using a polygon mesh based on the
\dune[PolygonGrid] \cite{dune:polygongrid}
with elements consisting on Voronoi cells around random
points - the methods \pyth{pyevolve} and \pyth{initialize} are not included
in the presentation.

\renewcommand{\pweavecaption}%
{Finite volume scheme using a polygonal mesh based on a Voronoi tessellation}
\renewcommand{\pweavelabel}{lstFVPoly}

\begin{pweavecode}
from dune.grid import CommOp
from dune.polygongrid import polygonGrid, voronoiDomain

boundingBox = numpy.array([[0, 0], [1, 1]])
view = polygonGrid(voronoiDomain(1513, boundingBox, seed=1234))

@dune.grid.gridFunction(view)
def c0(x):
    return 1.0 if x.two_norm > 0.125 and x.two_norm < 0.5 else 0.0
class Bnd:
    def __init__(self,t):
        self.t = t
    def __call__(self,x):
        return c0(x - [self.t, self.t])

figure = pyplot.figure(figsize=(20,10))

mapper = view.mapper(lambda gt: gt.dim == view.dimension)
c = initialize(view,mapper,c0)

@dune.grid.gridFunction(view)
def solution(element,x):
    return c[mapper.index(element)]

figParams = {"colorbar":False,"xlim":(0,1),"ylim":(0,1),"gridLines":"white"}
pyplot.subplot(121)
solution.plot(figure=figure,**figParams)

t = 0.0
count = 0
start = time.time()
while t < 0.5:
    t += pyevolve(view, mapper, c, Bnd(t))

pyplot.subplot(122)
solution.plot(figure=figure,**figParams)
pyplot.show()
\end{pweavecode}
\begin{figure}[htpb]
\center
\includegraphics[width= 0.95\textwidth]{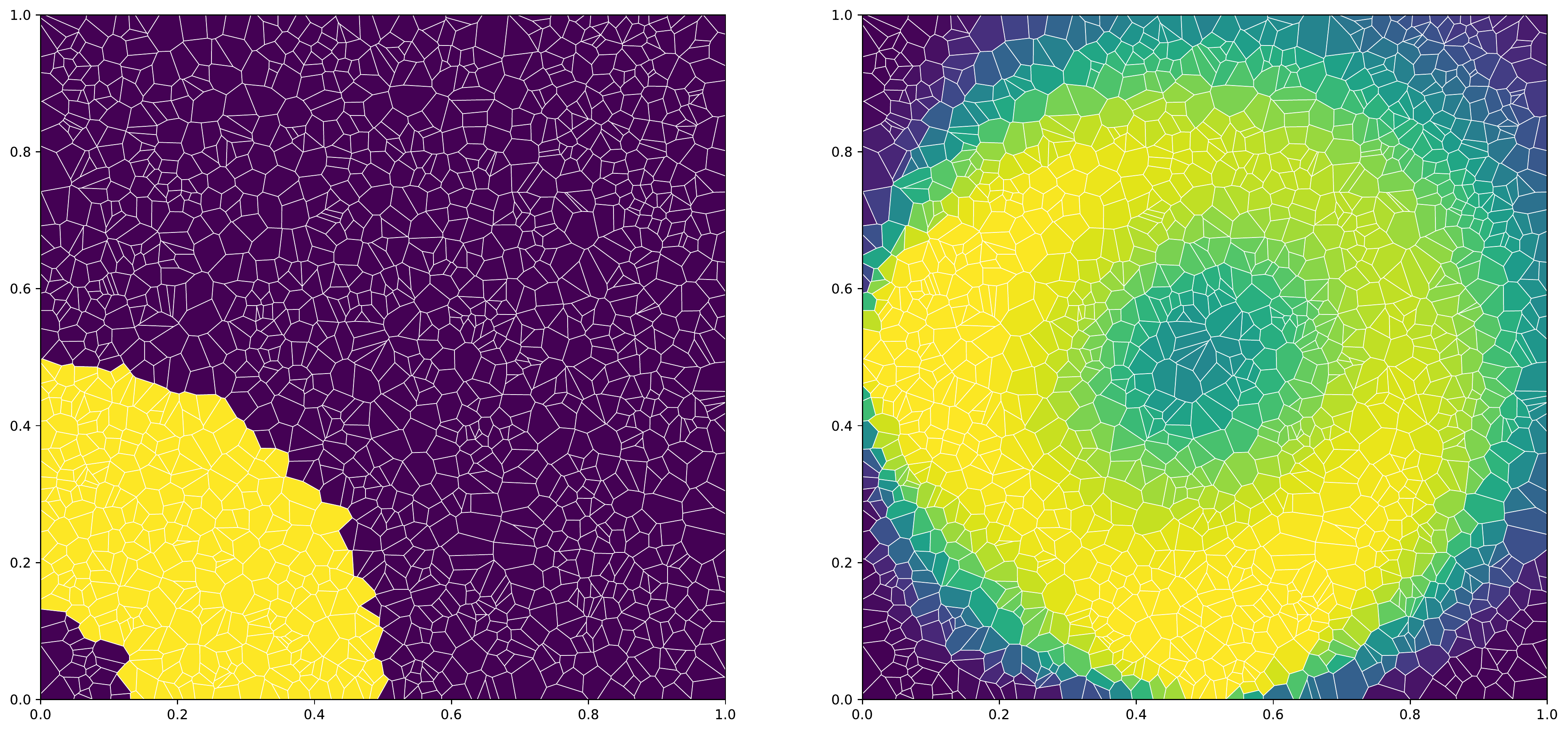}
\caption{Finite volume simulation: left initial conditions and right the solution at final time}
\label{fig:PyFVPoly}
\end{figure}

\begin{figure}[htpb]
\center
\includegraphics[width=0.47\textwidth]{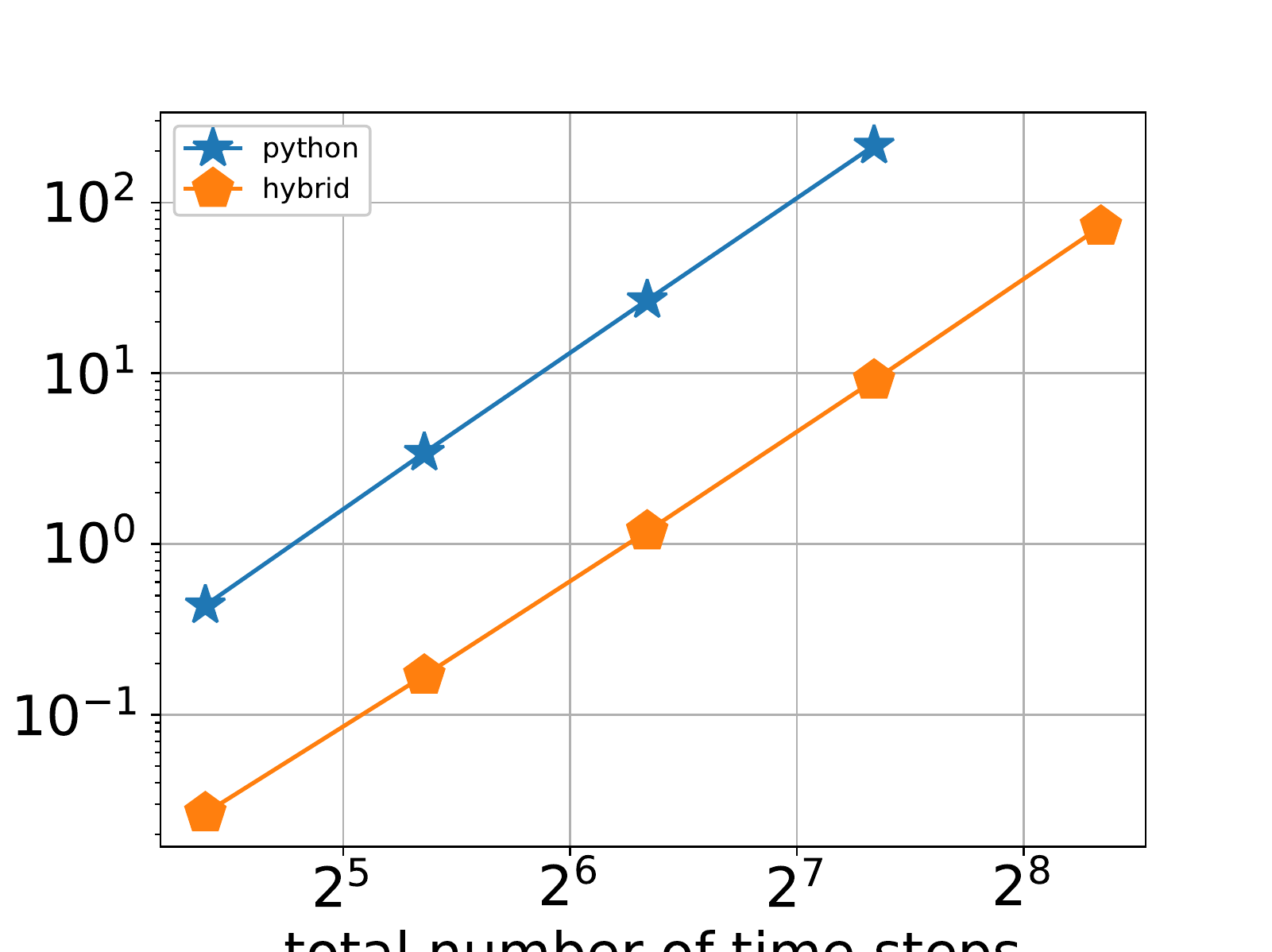}
\includegraphics[width=0.47\textwidth]{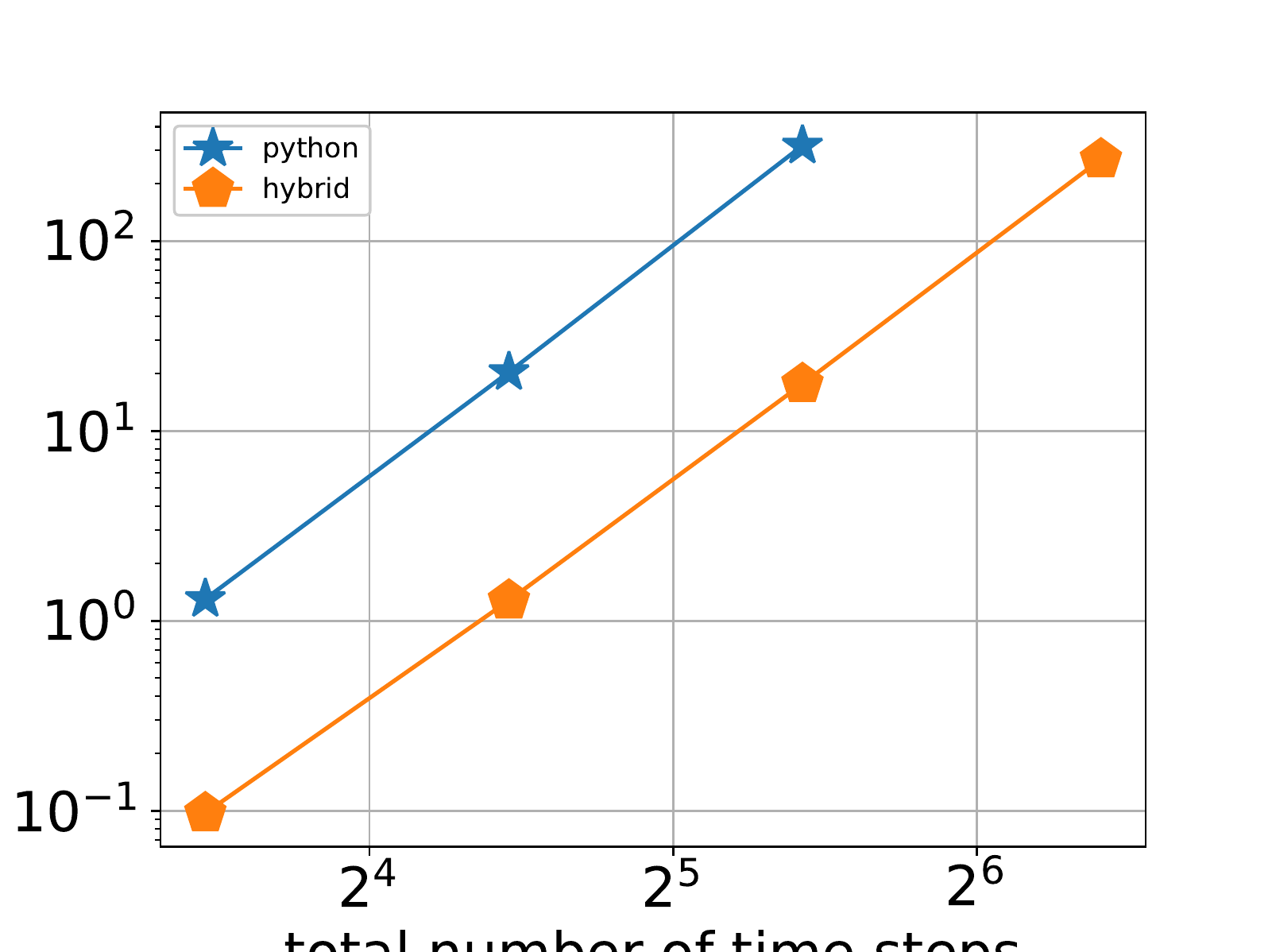}
\caption{Computational cost for pure Python and hybrid implementation of a
first order finite volume scheme using different levels of refienment of
a structured 2d \cpp{YaspGrid} on the left and an unstructured 3d \cpp{ALUGrid}
on the right.}
\label{fig:fvcompcost}
\end{figure}

To compare the runtime we use a version of the code based purely on the
Python interface similar to the one used above, as well as a version
where the grid based part of the scheme is rewritten in C++ (i.e., the
\pyth{pyevolve} method used above) and then
imported into Python using the \pyth{dune.generator.algorithm} function,
so that only the time loop, pre- and post-processing
are done in Python. The only callback in this version occurs at the
boundary to compute the boundary fluxes.
We compare the computational cost of both versions
using different grid implementations and resolutions.
Note that due to the simplicity of the finite volume method, vectorization
will not improve the speed of the Python code. In fact this is a very hard
problem for this type of interface since very little actual computation is
performed compared to the required number of calls to the grid interface.
Figure \ref{fig:fvcompcost} again shows about a factor of ten between the
pure Python and the hybrid Python/C++ implementations.

\section{Conclusions}

In this paper we presented \dune[Python], a new \dune module adding Python
interoperability to the \dune C++ environment.
This module provides Python bindings for the main classes in the \dune core
modules with a strong focus on the mature grid interface.

We achieved this by introducing a general concept for the on-the-fly generation
of binding code for statically polymorphic interfaces.
A small amount of C++ code using \pybind has to be written for each interface
class to be exported.
An even smaller amount of Python code has to be provided for each implementation
of an exported interface.
These functions are needed to convert dynamic parameters to template arguments
to generate the full C++ type.
The concept also allows to add realization specific extensions, such as
additional methods or different constructors.
One important underlying concept is a type registry associating each exported
C++ class to the C++ type and its required includes.
This allows us to easily instantiate complex C++ class templates on the Python
side.
This concept for binding statically polymorphic C++ interfaces to dynamically
typed languages is very general and not restricted to the Python language or to
the \dune environment.

In the second part of the paper we demonstrated how prototypes for
complex numerical schemes can be implemented using the provided bindings
for the \dune core module.
It is possible to write Python code which is very close to its C++ counterpart,
making it straightforward to translate the scheme into highly efficient C++
code.
This allows for rapid prototyping with the aim of using the C++ interface
for production code at a later stage.

We also introduced concepts for improving the efficiency of the Python code by
deviating slightly from the C++ interface.
The key here is the reduction of calls between Python and C++ and, hence,
vectorization support was added to some critical parts of the interface.
Using these features, small to middle sized problems can be solved relying solely
on the Python interface.

Problably the most useful approach, however, is the combination of both
techniques, i.e., implement the expensive functionality of the program in C++
functions but keep the problem-specific high-level code in Python.
Based on the type registry, this approach is supported by additional JIT
compilation methods to generate bindings for single C++ function templates.

\subsection*{Shortcomings and Outlook}

Some central functionality of the \dune core modules is not yet supported by
\dune[Python].
For example, bindings for the solvers and preconditioners implemented in the
\dune[Istl] module are not yet available and only rudimentary bindings for
\dune[LocalFunctions] have been included in the initial release.

While bindings for most of the interfaces in \dune[Grid] are part of
\dune[Python], some less-used concepts are still missing, e.g.,
local adaptivity with data transfer from the old grid hierarchy to the new
one.
Apart from exporting additional concepts, such as the \cpp{IdSet} or the
\cpp{PersistentContainer}, to Python, an efficient convenience interface
reducing both the required in-depth knowledge of \dune and the required number
of interlanguage calls, has to be designed.

Based on the concepts described in this paper, a number of further \dune modules
are in the process of providing Python bindings.
Complete bindings for \dune[ALUGrid] \cite{dune:alugrid}, \dune[SPGrid], and
\dune[PolygonGrid] are already available.
Preliminary bindings for the remaining core modules \dune[Istl] \cite{dune:istl}
and \dune[LocalFunctions] \cite{dune:localfunctions}, as well as the new
\dune[Functions] module, have been written in the context of \dune[Python]
itself.

To improve the usability and flexibility of our Python bindings for
\dune, more advanced code generation concepts need to be included.
At the time of writing we are developing the \dune[Fempy] module
\cite{dune:Fempy} exporing the interfaces in \dune[Fem] \cite{dune:Fem} to
Python.
It provides, for example, bindings for general finite element spaces, full hp
adaptivity, and commonly used solver packages, like PetSc.
These bindings are complemented by using the domain specific language UFL
\cite{ufl} to generate code for general grid functions as well as (bi-)linear
forms on discrete function spaces.
Consequently, the system matrix for a complex PDE problem can be assembled
without any callback into Python, thus eliminating one of the efficiency issues
described above.
An application to porous media flow is presented in \cite{twophasedg}.

\section*{Acknowledgements}

The authors would like to thank Micha{\"e}l Sgha{\"i}er for writing and testing
parts of the Python bindings for the \dune[Grid] interface during the Google
Summer of Code 2016, as well as Google Inc.\ for organizing and financing the
Summer of Code.


\bibliographystyle{abbrvnat}
\bibliography{dune-python}

\appendix

\section{Installation}
\label{sec:installation}

The simplest approach for testing \dune[Python] is to use the provided
Docker image. It includes a number of Jupyter notebooks showcasing
the possibilities of the \dune Python bindings. There is a special Docker
image accompanying this paper. It can be used by executing
\begin{bash}
docker run --rm -v dune-python-paper:/dune -p 127.0.0.1:8888:8888 registry.dune-project.org/staging/dune-python:paper2018
\end{bash}
The Jupyter server can then be accessed from a web browser at \url{http://127.0.0.1:8888};
the password is \file{dune}.
The code examples from this paper are included in the notebook file
\file{paper2018.ipynb}.
Notice that this method works on Linux, MacOS, and Windows alike, although it
might be necessary to increase the amount of system memory given to Docker to
4~GB, e.g., on MacOS.

If you already have the \dune core modules either installed or in local
space, it suffices to download the \dune[Python] module
\begin{bash}
export GITURL=https://gitlab.dune-project.org/staging/dune-python
wget -qO - ${GITURL}/repository/archive.tar.gz?ref=releases/2.6 | tar xz
\end{bash}
and to configure the module by running \code{dunecontrol}. Then
set the environment variable \code{PYTHONPATH} to include the \code{python}
subfolder in the build directory of \dune[Python], e.g.,
\code{dune-python/build-cmake/python}.

For some of the example \dune[ALUGrid] and for the final example
\dune[Polygongrid] will be required in addition to
the core modules. After building these modules the corresponding `python`
subfolder in the build directories of these modules also need to be added
to \code{PYTHONPATH}.
For a more permanent installation of \dune[Python] we suggest to set up a
virtual environment and install all necessary \dune modules into it.
More details can be found in the \url{README.md} file or on the main page of
the GitLab repository of \dune[Python].

%

\section{Changes to the C++ interface}
\label{sec:changedinterfaces}

The following give a short overview of changes and extensions we made
to the \dune\ interface while exporting it to Python. Some changes are
required due to the language restriction in Python or to make the
resulting interface more Pythonic.
Other changes were made, to make writing efficient code possible, e.g.,
vectorization, have already been discussed above.
Overall, this summary targets Dune developers familiar with the C++
interface to avoid unnecessary surprises.
However, it also provides guide-lines to export future interfaces
to Python.

\begin{itemize}
\item
  Since \pyth{global} is a keyword in Python we cannot export the
  \cpp{global} method on the \cpp{Geometry} directly. So we have
  exported \pyth{global} as \pyth{toGlobal} and for symmetry reasons 
  \pyth{local} as \pyth{toLocal}.
\item
  Some methods take compile-time static arguments, e.g., the codimension
  argument for
  \cpp{entity.subEntity< c >( i )}.
  These had to be turned into dynamic arguments, so in Python the subEntity
  is obtained via \pyth{entity.subEntity(i, c)}.
\item
  In many places we replaced methods with properties, i.e.,
  \pyth{entity.geometry} instead of \cpp{entity.geometry()}.
\item
  Methods returning a \pyth{bool} specifying that other interface methods
  will return valid results are not exported (e.g. \pyth{neighbor}
  on the intersection class). Instead \pyth{None} is returned to specify
  a non valid call (e.g. to \pyth{outside}).
\item
  Some of the C++ interfaces contain pairs of methods where the
  method with the \emph{plural name} returns an integer (the \emph{number of})
  and the singular version takes an integer and returns the \emph{ith}
  element.

  Consider, for example, \pyth{geometry.corners()} and \pyth{geometry.corner(i)}.
  Using the methods, loops would read as follows:
\renewcommand{\pweavecaption}{Standard C++ interface for iterating over corners of a geometry}
\renewcommand{\pweavelabel}{lstCppIterateCorners}
\begin{python}
for i in range(geometry.corners):
    print(geometry.corner(i))
\end{python}

  For Python users, this is very unintuitive and, even worse, it invokes a
  lot of expensive calls into the C++ code.
  For these reasons, we decided to slightly change the semantics of these methods:
  the plural version now returns a tuple or list object.
  The singular version still exists in its original form.
  So the above code snippet should be written as follows:
\renewcommand{\pweavecaption}{Python interface for iterating over corners of a geometry}
\renewcommand{\pweavelabel}{lstPyIterateCorners}
\begin{python}
for c in geometry.corners:
    print(c)
\end{python}
  Note that to obtain the original value returned by \cpp{geometry.corners}
  we now need to write \pyth{len(geometry.corners)}.
\item
  In C++, free-standing functions can be found via argument-dependent lookup.
  As Python does not have such a concept, we converted those free-standing
  functions to methods or properties.
  Examples are \cpp{elements}, \cpp{entities}, \cpp{intersections}, or
  \cpp{localFunction}.
\item
  A \emph{grid} in \dune[Python] is always the \cpp{LeafGridView} of the
  hierarchical grid. To work with the actual hierarchy, i.e., to refine the
  grid, use the \pyth{hierarchicalGrid} property. Level grid view can
  also be obtained from that hierarchical grid.
\item
  In contrast to C++, partitions are exported as objects of their own.
  The interior partition, for example, can be accessed by
\renewcommand{\pweavecaption}{Python interface for iterating over all elements of a grid}
\renewcommand{\pweavelabel}{lstPyIterateElements}
\begin{python}
partition = grid.interiorPartition
\end{python}
  The partition, in turn, also exports the method \pyth{entities} and the properties
  \pyth{elements}, \pyth{facets}, \pyth{edges}, and \pyth{vertices}.
\item
  A \pyth{MCMGMapper} can be constructed using the \pyth{mapper}
  method on the \pyth{GridView} class passing in the \pyth{Layout}
  as argument.
  The mapper class has an additional call method taking an entity,
  which returns an array with the indices of all dofs attached to that
  entity.
  A list of dof vectors based on the same mapper can be communicated using
  methods defined on the mapper itself and without having to define a
  \cpp{DataHandle}.
\end{itemize}

\end{document}